\documentclass[aps, nofootinbib,  preprint, 12pt]{revtex4}
\usepackage{graphicx}

\newcommand{\be}{\begin{equation}}
\newcommand{\ee}{\end{equation}}
\newcommand{\bea}{\begin{eqnarray}}
\newcommand{\eea}{\end{eqnarray}}

\newcommand{\rpbp}{\bar{p}/p}

\def\urltilde{\lower .7ex\hbox{\~{}}}

\def\lsim{\mathrel{\rlap{\lower4pt\hbox{\hskip1pt$\sim$}}
    \raise1pt\hbox{$<$}}}         
\def\gsim{\mathrel{\rlap{\lower4pt\hbox{\hskip1pt$\sim$}}
    \raise1pt\hbox{$>$}}}         

\usepackage{epsf,epsfig,subfigure,graphicx,amsmath,amssymb}
\usepackage{color}

\begin{document}
\draft
 \begin{flushright}
{\tt IPMU08-0023 \\
      YITP-09-16\\
      KEK-TH-1306}
\end{flushright}

\title{\Large Dark matter and collider phenomenology of split-UED}


\author{Chuan-Ren Chen${}^1$, Mihoko M. Nojiri${}^{123}$, Seong Chan Park${}^1$, Jing Shu${}^1$, Michihisa Takeuchi${}^{24}$} 
\vspace{1cm}
\affiliation{\small ${}^1$Institute for the Physics and Mathematics of the Universe, The University of Tokyo, 
Chiba $277-8568$, Japan\\
${}^2$Theory Group, KEK, 1-1 Oho, Tsukuba, Ibaraki 305-0801, Japan\\
${}^3$The Graduate University for Advanced Studies (SOKENDAI),1-1 Oho, Tsukuba, Ibaraki 305-0801, Japan\\ ${}^4$Yukawa Institute for Theoretical Physics, Kyoto University,
 Kyoto 606-8502, Japan
}

\vspace{1.0cm}

\begin{abstract}
We explicitly show that split-universal extra dimension (split-UED), a recently suggested extension of universal extra dimension (UED) model, can nicely explain recent anomalies in cosmic-ray positrons and electrons observed by  PAMELA  and ATIC/PPB-BETS.  Kaluza-Klein (KK)  dark matters mainly annihilate into leptons because the hadronic  branching fraction is highly suppressed by large KK quark masses and the antiproton flux agrees very well with the observation where no excess is found . The flux of cosmic gamma-rays from pion decay is also highly suppressed and hardly detected  in low energy region ($E_\gamma\lesssim 20$ GeV). Collider signatures of colored KK particles at the LHC, especially $q_1 q_1$ production, are studied  in detail. Due to the large split in  masses  of KK quarks and other particles, hard $p_T$ jets and missing $E_T$ are generated, which make it possible to suppress the standard model background and discover the signals.
 \end{abstract}


\maketitle

\section{Introduction}

Recent observations by PAMELA \cite{PAMELA} and ATIC/PPB-BETS \cite{ATIC, PPB-BETS} consistently suggest a new primary source of energetic cosmic electrons and positrons in the neighborhood of our solar system, since high energy electrons/positrons lose their energy quickly within about 1 kpc.
These resutls have attracted  enormous attentions and stimulated many interpretations of the primary electric source from both astrophysics and particle physics, for example, the nearby pulsars \cite{pulsar} and dark matter decay \cite{decayDM} or annihilation \cite{anniDM}.
The existing data of cosmic electrons and positrons cannot distinguish among these possibilities, and we expect that future data from cosmic-ray and accelerator experiments such as Fermi and Large Hadron Collider (LHC) will give us some clues on which interpretation is more promising and ultimately correct. 

In this paper, we consider one of the most attractive models of dark matter, universal extra dimension (UED)  with its minimal realization (mUED)\cite{UED,Cheng:2002ab} and a recently suggested variety split-UED \cite{split-UED}. When the reflection symmetry (dubbed Kaluza-Klein parity) about the mid point of the orbifold extra dimension is exact, the lightest Kaluza-Klein particle (LKP), which is odd under the parity operation, is absolutely stable and is a good candidate of dark matter \cite{Servant:2002aq}.
It turns out that in most cases of universal extra dimension scenarios the LKP is the first Kaluza-Klein (KK) photon ($B_1$), and a pair of charged leptons is produced in the annihilation process of $B_1$. Interestingly enough, the mass of dark matter suggested by the  ATIC/PPB-BETS data ($\sim 600-700$ GeV) exactly coincides with the one that  gives the right relic density for the LKP dark matter after taking co-annihilation channels into account \cite{Kong:2005hn}.

On the other hand, the measured data of the cosmic ray antiproton-to-proton flux ratio between 1 and 100 GeV by PAMELA follow the expectations from secondary production and strongly constrain contributions from dark matter particle annihilation \cite{Adriani:2008zq}.  Therefore it is required to naturally reduce antiproton flux in any dark matter model.  Split-UED is promising in the sense that production of antiproton is quite suppressed. The cross section for dark matter annihilates into a quark pair is 
$$\langle \sigma_{B_1B_1\to q \bar{q}} v \rangle \propto m_{B_1}^2/(m_{B_1}^2+m_{q_1}^2)^2$$ 
where $m_{B_1}$ and $m_{q_1}$ are the masses of $B_1$ and the first  KK quark ($q_1$), respectively, and it can be significantly suppressed by increasing the masses of KK quarks, i.e. splitting the KK quarks from other particles. However we should keep in mind that predictions of the observed antiprotons from dark matter annihilation can vary quite a lot for different adopted diffusion models as we will see later. As hadronic production is suppressed in split-UED, we would also expect much less production of cosmic gamma-ray whose main source is decay of pions.  This prediction will be tested by forthcoming data from the Fermi experiment \cite{FGST}.

The split spectrum of Kaluza-Klein particles is realized by introducing a bulk mass term for quarks in split-UED keeping the KK parity exact \cite{split-UED} 
\footnote{In warped geometry KK parity could be kept in a different way \cite{Agashe:2007jb}.}. 
One immediate consequence is that the 5D wave functions of quark fields are quasi-localized at the boundaries and induce violation of the translational symmetry along the extra dimension. Accordingly the KK number conservation is violated in the quark sector, and KK-even gauge bosons could directly couple to the zero mode quarks at the tree level, which is forbidden in mUED. KK-even bosons could be copiously produced at colliders and it is promising to discover them, for example, by the 2nd KK gluon resonance  at the LHC \cite{split-UED}. Another consequence is the heavy KK quark decay to the SM quark and KK boson, which will generate  harder jets with large missing momentum compared to that in mUED due to a large mass splitting between KK quarks and other KK particles. Such differences may give us some handle to tell split-UED apart from mUED at the LHC.

This paper is organized as follows. In Section \ref{sec:model} we review split-UED and show how the spectrum and the interactions are modified compared with those in mUED.  Note that the renormalization group (RG) running is taken into account in  our spectrum  calculation. We then calculate the annihilation cross section to leptons and quarks with varying ``bulk mass parameter", $\mu$, which is the only new extra parameter in split-UED. 
In Section \ref{sec:cosmic-ray}, we estimate the signals of cosmic-ray positrons, electrons, and antiprotons from LKP dark matter annihilation in both mUED \cite{Hooper:2009fj} and split-UED, and compare our predictions with recent experimental data. Moreover, the diffusive gamma-ray is studied as well. In Section  \ref{sec:collider}, we  consider the productions of KK quarks and gluons at the LHC, and we also discuss the possibility of discovering  these signals and mass reconstructions. We finally conclude our studies in Section \ref{sec:conclusion}.

\section{Masses and Couplings in Split-UED}
\label{sec:model}

Our set-up is based on the all SM fields are ``universally" propagating in one extra dimension, which is compactified on an orbifold, $S^1/Z_2$, with the boundary points $y = \pm L$ (the orbifold radius $R = 2L/\pi$). The extension we have introduced is an scalar background $\Phi(y)$ that couples to each colored Dirac fermions $\Psi_{q}(x^\mu,y)$ in the 5D bulk where the corresponding standard model quark field $q$ (=$SU(2)$ singlets $u_R^c, d_R^c$ and doublet $Q_L$) resides on the chiral zero mode after orbifold projection.  A common Yukawa coupling $\lambda(\equiv\lambda_{Q_L}=\lambda_{u_R^c}=\lambda_{d_R^c})$ is assumed for {\it all} quarks so that any further source of flavor violation beyond the CKM matrix is automatically avoided and the model becomes more predictive \footnote{In general the Yukawa coupling can be assigned for any fermion field in the bulk and they can be different from each other.}. Here we neglect the family indices but one can easily extend the model to include more generations. Finallly the 5D action is given by the form
 \begin{eqnarray}
 S_{\rm split-UED}=\int d^4 x \int_{-L}^{L} dy \Big[\mathcal{L_{\rm mUED}} - \lambda \Phi(y) \bar{\Psi}_q(x,y) \Psi_q(x,y)  \Big] \ ,
 \label{eq:5D}
\end{eqnarray}
where the summation runs for all quark fields $q$, and $\mathcal{L_{\rm mUED}}$ contains the usual mUED terms. 

      If we choose an odd background profile $ \Phi(-y) = - \Phi(y)$ under the inversion symmetry about the middle point $y =0$, we can see that the KK parity, which transforms the fields as
$\Phi(x,y) \rightarrow \Phi(x,-y)$, $\Psi_q(x, y) \rightarrow \pm \gamma_5 \Psi_q(x, -y) $, is  still a good symmetry of the Lagrangian. For quantitative concreteness we consider the simplest mass profile which respects the requirement: $\lambda \langle \Phi(y) \rangle = \mu \epsilon(y)$, where $\mu$ is the bulk mass parameter and $\epsilon(y)$ is a step function defined as $+1$ for $0<y<L$ and $-1$ for $-L<y<0$. As ($\lambda\to 0$) or equivalently ($\mu \to 0$), split-UED is continuously reduced to mUED thus we call this limit as ``mUED limit" below. For gauge bosons, scalars, and leptons ($l$), their KK decompositions are the same as those in mUED, while for quarks, their odd masses will affect the KK decompositions. The details of the all KK decompositions in Eq. (\ref{eq:5D}) are presented in Appendix A. Notice that in the case of KK decomposition, we also label the KK parity in addition to the KK number $ n = \{ n^- (\textrm{n odd}), ~n^+ (\textrm{n even}) \}$.

The tree level mass for the $n$-th KK fermion is given by
\begin{eqnarray}
m_{q_n}^\textrm{tree} &=& \sqrt{\mu^2 + k_n^2}, \\
m_{l_n}^\textrm{tree} &=& n/R \ ,
\end{eqnarray}
if we neglect their masses coming from electroweak symmetry breaking.
Here $k_n$ is determined by the boundary conditions which are given by: 
\begin{eqnarray}
k_{n^-} &=& - |\mu| \tan{k_n L}, \\
k_{n^+} &=& n/R \ .
\end{eqnarray}
We choose the minimal universal boundary conditions at the energy cutoff scale $\Lambda$ that all boundary kinetic terms vanish \footnote{One can also consider the non-universal boundary conditions as an extension. See for example, Ref. \cite{Flacke:2008ne}. }, and the general expression for the one-loop correction to the $n$-th KK fermion  masse is given by \cite{Cheng:2002iz}
\begin{eqnarray}
m_{f_n}=m_{f_n}^\textrm{tree}\left[1+ \frac{9}{8\pi}\left( \sum_G C_2^G(N) \alpha_G \right)\log \Lambda R\right] \ ,
\end{eqnarray}
where $\alpha_G=g_G^2/4\pi$ and we neglect the Yukawa corrections in this formula\footnote{For the KK top quark, if we consider the large top Yukawa coupling, positive contribution to the tree level mass term will cancel the negative one from the radiative corrections, and the overall mass corrections will not be affected too much.}. The sum is over all SM gauge groups under which the fermion $f$ is charged, and $C_2^G(N)$ is the quadratic Casimir operator of the fundamental representation $N$ in gauge group $G$, e.g., $C_2^{SU(N)}(N)=(N^2-1)/2N$ and we take $C_2^{U(1)_Y}(N)=Y_f^2$ with $Y_f$ being the hypercharge of $f$. If we choose $(\alpha_1,\alpha_2,\alpha_3)\simeq(0.010, 0.033, 0.094)$ at $Q^2=(620 {\rm GeV})^2 $,  colored particles will get bigger corrections ($\sim 14\%$ for $u_R^1, d_R^1$ and $16\%$ for $Q_L^1$) than leptons ($\sim 1.9\%$ for $l_R^1$ and $3.1\%$ for $l_L^1$). 

\begin{figure}[t]
\includegraphics[width=.6\textwidth, angle=0]{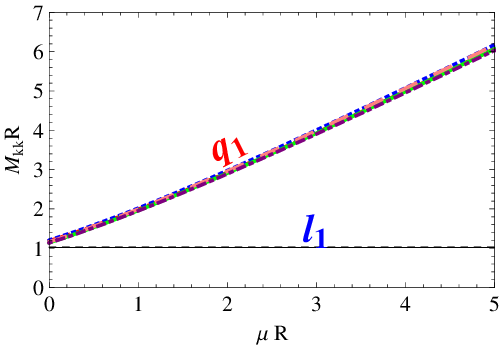}
\includegraphics[width=.63\textwidth, angle=0]{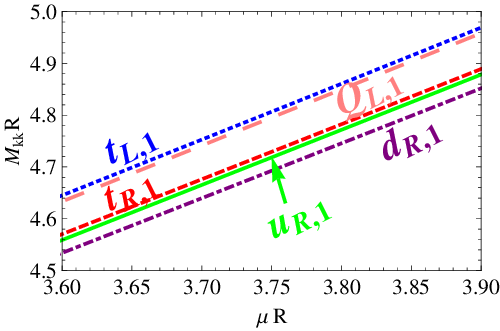}
\caption{\label{fig1} (upper plot) The spectra of 1st KK quarks and leptons in the unit $1/R$ with the given bulk mass parameter $\mu$. The lepton masses are independent of $\mu$ in our scenario.
Here we included 1-loop correction for all the masses. The most important correction is QCD correction
and it is roughly $14\sim16\%$ for quarks. For leptons the correction is small ($\sim 2\%$).
(lower plot) The magnified figure of the upper one in $\mu R = (3.60, 3.90)$. $SU(2)$ doublets ($t_L, Q_L$) get the larger 1-loop correction than singlet quarks ($u_R, d_R$) get.
}
\end{figure}
In Fig. \ref{fig1} we plot the spectra of 1st KK quarks and leptons including 1-loop corrections. When $\mu=0$, i.e. the mUED limit, quarks are quite degenerate with leptons even we take the QCD corrections to the masses into account. However when we increase the bulk mass parameter $\mu$, the KK quark masses become larger and larger so that we get the split spectra as we request. Note that small deviations in quark spectra come from different $SU(2)$ and $U(1)$ charge of $u_R, d_R$ and $Q_L$. In our study, these differences can be neglected.
\begin{figure}[t]
\includegraphics[width=0.75 \textwidth, angle=0]{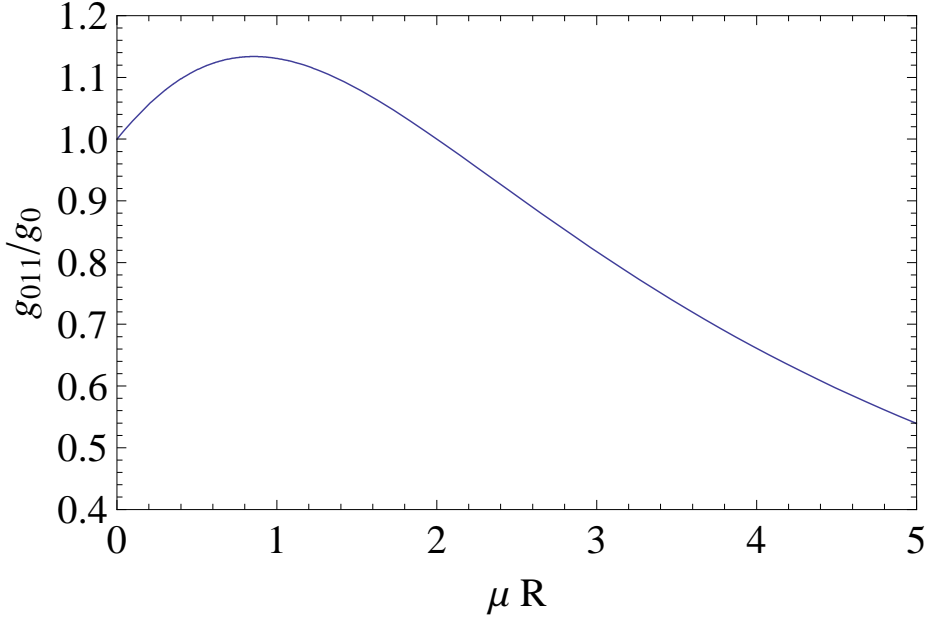}  
\caption{\label{fig2} The coupling constant ratio $g_{A_1-q_1-q_0}/g_0 =  {\cal G}_{110}$ vesus $\mu R$. The dependence comes from the change of the 5D profiles. Notice that the coupling ratio first increases to its maximal value around 1.14 for $\mu R \sim 1$ and then decreases monotonically.  }
\end{figure}

The couplings between KK gauge bosons and fermions determine the most interesting phenomenological features of the model. All 4D effective couplings could be calculated by integrating out the 5D profiles of the relevant fields along the extra dimension. Since 5D profiles are the same for different KK gauge bosons (quarks) with the same KK number, the ratio between a general gauge boson-quark-quark  coupling $g_{A_m-q_n-q'_l}$ over the SM one is the same for fixed KK numbers, where $m$, $n$ and $l$ are KK number of $A$, $q$ and $q'$, respectively. We introduce a function $\mathcal{G}_{mnl} (\mu R)$ to parameterize how such a ratio depends on the bulk mass parameter $\mu$ times the radius $R$:  
\bea
g_{A_m-q_n-q_l}/g_0 \equiv  {\cal G}_{mnl}(\mu R) \ ,
\eea 
where $g_0$ is the SM gauge interaction between $A_0$, $q_0$ and $q'_0$. 
Among those couplings, the most interesting one in phenomenology is the coupling between the 1st KK gauge boson $A_1$, the 1st KK quark $q_1$, and SM quark $q_0$, which is crucial on LKP pair annihilation and productions of $q_1$ and the 1st KK gluon $g_1$ at the LHC. 
The dimensionless function ${\cal G}_{110} (\mu R)$ is calculated from the 5D profiles of $A_1$, $q_1$ and $q_0$ given by Eq. (\ref{eq:even}), (\ref{eq:odd}), (\ref{eq:quarkKK}), and (\ref{eq:quarkzero}) in Appendix A, 
and is plotted in Fig. \ref{fig2} as a function of $\mu R$. 
At $\mu=0$ limit we get the mUED result, which is the same as the SM coupling. As $\mu$ gets larger, the 5D profiles are localized so that the resultant overlap among the fields get smaller.   

\begin{figure}[t]
\includegraphics[width=0.75 \textwidth, angle=0]{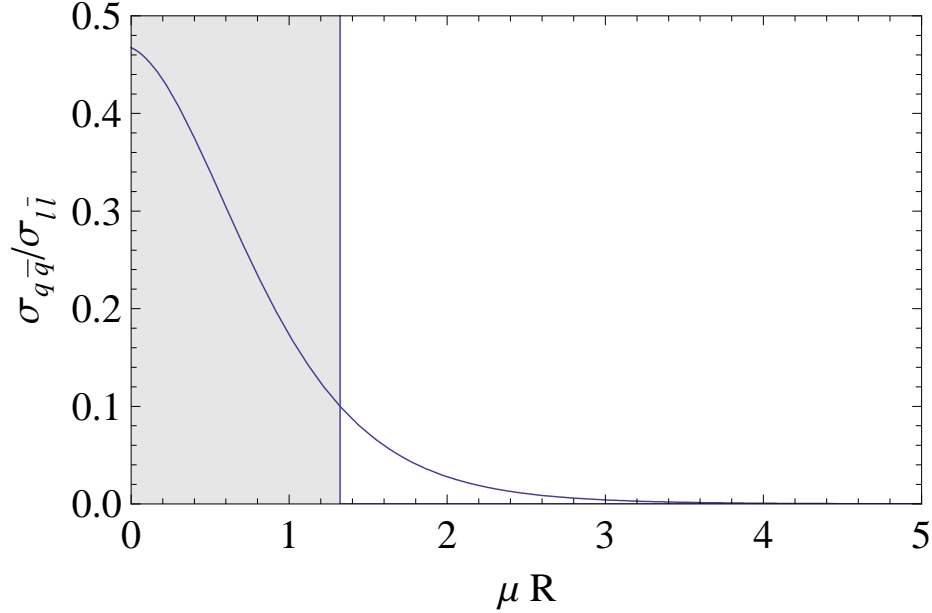}  
\caption{\label{fig3} The ratio of LKP pair annihilation into quarks and charged leptons vesus $\mu R$. The mUED limit corresponds to $\mu=0$ and the ratio of cross section to quarks to  leptons is about $47\%$ but it quickly goes down below $10\%$ as $\mu \gsim 1.3/R$. The shaded region ($\sigma_{q\bar{q}}/\sigma{l\bar{l}} > 10\%$) is not preferred by PAMELA antiproton data.}
\end{figure}

\begin{table}[]
\caption{\label{table1} The branching fraction of the LKP pair annihilates into various final states, for different $q_{1}$ mass, and different bulk mass parameter $\mu$. The numbers shown are summed over generations. We neglect the $W$-boson final state because the SU(2) component of LKP is too small. Here we fix the LKP mass to be 620 GeV and include the radiative corrections to all the masses according to Ref. \cite{Cheng:2002iz}.}
\begin{ruledtabular}
\begin{tabular}[c]{c|c|c|c|c|c|c}
$\mu$ (GeV)& 0   & 200  & 400  & 600 & 800 & 1000\\
   \hline
   $M_{q_{1}}$ (GeV)& 713   & 863  & 1026 & 1198 & 1378 & 1566\\
   \hline
    \textrm{BR}($B_1 B_1 \rightarrow q\bar{q}$) & 29.4\% &  26.4\%  &  20.6\% &  14.3\% & 8.9\% & 5.2\%    \\ \hline
        \textrm{BR}($B_1 B_1 \rightarrow l\bar{l}$) & 64.3\% &  67.1\%  &  72.3\% &  78.2\% & 83.0\% & 86.5\%    \\ \hline
            \textrm{BR}($B_1 B_1 \rightarrow \nu \bar{\nu}$) & 3.8\% &  3.9\%  &  4.3\% &  4.6 \% & 4.9\% & 5.1\%    \\ \hline
\textrm{BR}($B_1 B_1 \rightarrow \phi \phi^*$) & 2.3\% &  2.4\%  &  2.6\% &  2.8\% & 3.0\% & 3.1\%
\end{tabular}
\end{ruledtabular}
\end{table}

Taking care of all the corrections and effects of $\mu$ we can now calculate the branching fractions that LKP pair annihilates into quarks and leptons. At the non-relativistic limit ($\beta =\sqrt{1-4m_{B_1}^2/s}\rightarrow 0$), the cross section for LKP pair annihilates into fermions can be well approximated as:
\begin{eqnarray}
\langle\sigma v\rangle_{B_1 B_1\rightarrow f \bar{f}} =
\frac{2 g_1^4 C_f }{9\pi m_{B_1}^2} \left[ \frac{\mathcal{G}_{110}(\mu R)}{(1+r_f^2)} \right]^2 +{\cal O}(v), 
\end{eqnarray}
where a useful numerical term $C_f=N_c (Y_{f_L}^4+Y_{f_R}^4)$ is introduced with the color factor $N_c=3(1)$ for quarks(leptons). The mass ratio between the LKP and KK fermion is parameterized by a parameter $r_f$ defined as $r_f \equiv m_{f_1}/m_{B_1}$ and we ignore the small mass corrections coming from $SU(2)_L$, $U(1)_Y$ gauge group so that $m_{f_1}=m_{f_{L1}}\simeq m_{f_{R1}}$. In mUED limit, the ratio between productions of the lepton  and the quark is mainly controlled by the ratio of $C_f$ from the hypercharges and color factor of fermions. For leptons and quarks the corresponding terms are $\sum_l C_l \simeq 3.18 $ and $\sum_q C_q \simeq 1.90$, respectively. Obviously the mUED does have a sizable hadronic annihilation channel and a carefully study of antiproton flux is important. In the case of split-UED, one can easily adjust $r_q$ to suppress the hadronic annihilation cross section. In Fig. \ref{fig3} we plot the ratio of annihilation cross section into quarks and charged leptons $\sigma_{q\bar{q}}/\sigma_{l\bar{l}}$ with $\mu R$ in the range of $(0,5)$. We find the ratio goes down very quickly as the bulk mass parameter goes larger since  $\sigma \sim 1/(1+r_f^2)^2$. By turning on the bulk mass parameter we can easily avoid the antiproton excess in the cosmic ray detection experiments. We also note that  the soft gamma-ray from decay of hadrons, especially from pions, is under control in this set-up.

\section{Cosmic Positron, Antiproton and Photon}
\label{sec:cosmic-ray}

 LKP dark matter $B_1$ will mainly annihilate into a fermion pair while the W-boson and Higgs boson modes are suppressed, as shown in Table \ref{table1}. For the quark-pair final state, due to the QCD hadronization process, a bunch of hadrons will be produced and sequentially decay into positrons/electrons, protons/antiprotons, photons and neutrinos. Moreover, positrons and electrons  also originate in the leptonic channels: muon and tau (also generate photons) decay and direct production in $B_1 B_1 \to e^+ e^-$ process. In our calculation, we use PYTHIA \cite{Sjostrand:2006za} for the simulation of the QCD hadronization and the decays of particles. Since these stable particles from the dark matter annihilation in our Galaxy will propagate to our solar system and be observed as  cosmic-ray signals, we estimate the predictions from our model, the split-UED, and compared with experimental data in this section. Due to the fact that excesses were observed in the positron fraction in PAMELA \cite{PAMELA}  and the flux of electron plus positron in ATIC/PPB-BETS \cite{ATIC,PPB-BETS} while the  ratio of antiproton to proton, $\rpbp$, is consistent with the astrophysical background, we first explain positron and electron data from the dark matter annihilation. Then  base on the parameters set by fitting the ATIC/PPB-BETS and PAMELA data, we calculate the $\bar{p}/p$
and photon flux. Since the W-boson and the Higgs boson productions are highly suppressed in UED models we are considering, we will not take into account their contributions in the following calculations. 

\subsection{Positron and Electron}
\label{subsec:ep}
\begin{table}
\begin{tabular}{|c|c|c|c|}
\hline 
Model&
$\delta$&
$K_0 (\rm{kpc^2/Myr})$&
$L (\rm{kpc})$\tabularnewline
\hline
\hline 
M2&
$0.55$&
$0.00595$&
$1$\tabularnewline
\hline
MED&
$0.70$&
$0.0112$&
$4$\tabularnewline
\hline
M1&
$0.46$&
$0.0765$&
$15$\tabularnewline
\hline
\end{tabular}
\caption{The parameters in the diffusion models which are comparible with B/C and generate minimum (M2), median (MED) and maximal (M1) positron flux. }
\label{tab:model_ep}
\end{table}
In this section, we calculate the cosmic-ray positrons and electrons from the $B_1 B_1$ annihilation which account for the PAMELA and ATIC/PPB-BETS data. Since the dark matter annihilation generates the same amount of electrons and positrons, we only show the formula for positron flux below. After being produced from the dark matter annihilation process, the positrons will propagate
in the magnetic field of the Milky Way. Because the magnetic fields are tangled,
the motion of the positron can be described by a diffusion equation.
Neglecting the convection and annihilation in the disk, the steady
state solution must satisfy
\begin{equation}
K(E)\bigtriangledown^2 f_{e^{+}}(E,\vec{r})+\frac{\partial}{\partial E}\left[b(E)f_{e^{+}}(E,\vec{r})\right]+Q(E,\vec{r})=0,\,
\label{eq:e_prop}
\end{equation}
where $f_{e^{+}}$ is the number density of $e^{+}$ per unit kinetic
energy, $K(E)=K_0 (E/\rm{GeV})^\delta$ is the diffusion coefficient, $b(E)=10^{-16}(E/\rm{GeV})^2$
is the rate of energy loss and $Q(E,\vec{r})$ is the source of producing
$e^{+}$. In the $B_1$ dark matter annihilation case,
\begin{equation}
Q(E,\vec{r})=\frac{1}{2}\left(\frac{\rho(\vec{r})}{m_{B_1}}\right)^2 \sum_i  \langle \sigma v \rangle_i \left(\frac{dN(E)_{e^+}}{dE}\right)_i,
\end{equation}
 where $dN_{e^{+}}/dE$ is the energy spectrum of $e^{+}$  
 obtained by using PYTHIA, the index $i$ runs over all quark and charged lepton pairs and $\rho(\vec{r})$ is the dark matter profile. In our numerical calculations, we adopt an overall boost factor $B_F$ and the isothermal halo model \cite{Bergstrom:1997fj} which is given as \footnote{Effectively we are using $\rho^2(r) = B_F \rho^2 _{halo}(r)$}
\be
\rho_{\textrm{halo}} (\vec{r})=\frac{\rho_0}{1+(r/r_c)^2},
\ee
where $r=|\vec{r}|$  is the distance from our Galactic center, $r_c=3.5$ kpc and $\rho_0$ is the parameter that is adjusted to yield a dark matter local halo density of $0.3\, \rm{GeV}/cm^3$ \cite{Bergstrom:1997fj}
in our solar system.
 Considering the diffusion zone of a cylinder with radius R and half-height L, the solution for the positron flux at the Earth can be written in a useful form \cite{Hisano:2005ec}

\begin{equation}
\Phi^{DM}_{e^{+}}(E)=\frac{c}{4\pi}f_{e^{+}}(E,r_{\odot}),\,
\label{eq:flux_e}
\end{equation}
where $c$ is the speed of light, 
\be
f_{e^{+}}(E,r_{\odot})=\int_{E}^{m_{B_1}}\frac{10^{16}}{E^2}dE'\sum_{n,m =1}^\infty Q_{n,m}(E')J_0\left(\frac{\zeta_n r_\odot}{R}\right)\sin\left(\frac{m\pi}{2}\right)\exp\left[ \left( (\frac{m\pi}{2L})^2+(\frac{\zeta_n}{R})^2\right)X\right]
\ee
where $E^{(')}$ is the energy in a unit of GeV, $J_0$ is the zeroth-order of the first kind Bessel function, $r_\odot\sim8.5$ kpc is the distance from Milky Way center to the Sun, $\zeta_n$ is the n-th zero of the function $J_0$,
 
\be
X=\frac{K_0\times10^{16}}{\delta -1}\left[ E^{\delta-1} -E'^{\delta-1}\right] 
\ee
and 
\be
Q_{n,m}(E) = \frac{2}{J_1(\zeta_n)^2 R^2 L}\int_0^R dr r J_0(\frac{\zeta_n r}{R})\int_{-L}^L dz \sin\left(\frac{m\pi z}{2L}(L-z)\right)Q(E,\vec{r}).
\label{eq:Rnm_e}
\ee
The $J_1$ in the Eq.(\ref{eq:Rnm_e}) is the first-order of the Bessel function and  the parameters $\delta, K_0$ are set in the simulations of diffusion models to coincide with the observed cosmic-ray data, especially the Boron to Carbon ratio (B/C) \cite{Maurin:2001sj}. The values for different diffusion models we adopted are listed in table \ref{tab:model_ep} \cite{Delahaye:2007fr}.

In addition to $e^{\pm}$ fluxes from dark matter annihilation, there exist
 secondary $e^{\pm}$ fluxes from interactions between cosmic-rays and
nuclei in the interstellar medium. We use the approximations of the $e^{-}$ and $e^{+}$
background fluxes \ \cite{Moskalenko:1997gh,Baltz:1998xv} 
\begin{align}
\Phi_{e^{-}}^{prim}(E) & =\frac{0.16E^{-1.1}}{1+11E^{0.9}+3.2E^{2.15}}\quad{\rm GeV}^{-1}{\rm cm}^{-2}{\rm sec}^{-1}{\rm sr}^{-1},\nonumber \\
\Phi_{e^{-}}^{sec}(E) & =\frac{0.7E^{0.7}}{1+110E^{1.5}+600E^{2.9}+580E^{4.2}}\quad{\rm GeV}^{-1}{\rm cm}^{-2}{\rm sec}^{-1}{\rm sr}^{-1},\nonumber \\
\Phi_{e^{+}}^{sec} & (E)=\frac{4.5E^{0.7}}{1+650E^{2.3}+1500E^{4.2}}\quad{\rm GeV}^{-1}{\rm cm}^{-2}{\rm sec}^{-1}{\rm sr}^{-1},\,
\label{eq:bg_e}
\end{align}
where $E$ is in units of GeV. Therefore, the fraction of positron
flux and total flux of  positron plus electron are
\begin{equation}
\frac{\Phi_{e+}^{DM}+\Phi_{e+}^{sec}}{\Phi_{e+}^{DM}+\Phi_{e-}^{DM}+\Phi_{e+}^{sec}+k\Phi_{e^{-}}^{prim}+\Phi_{e^{-}}^{sec}},\,
\label{eq:fract_e}
\end{equation}
and 
\be
\Phi_{e+}^{DM}+\Phi_{e-}^{DM}+\Phi_{e+}^{sec}+k\Phi_{e^{-}}^{prim}+\Phi_{e^{-}}^{sec},\,
\label{eq:flux_e}
\ee
\begin{figure}
\includegraphics[width=0.75\textwidth]{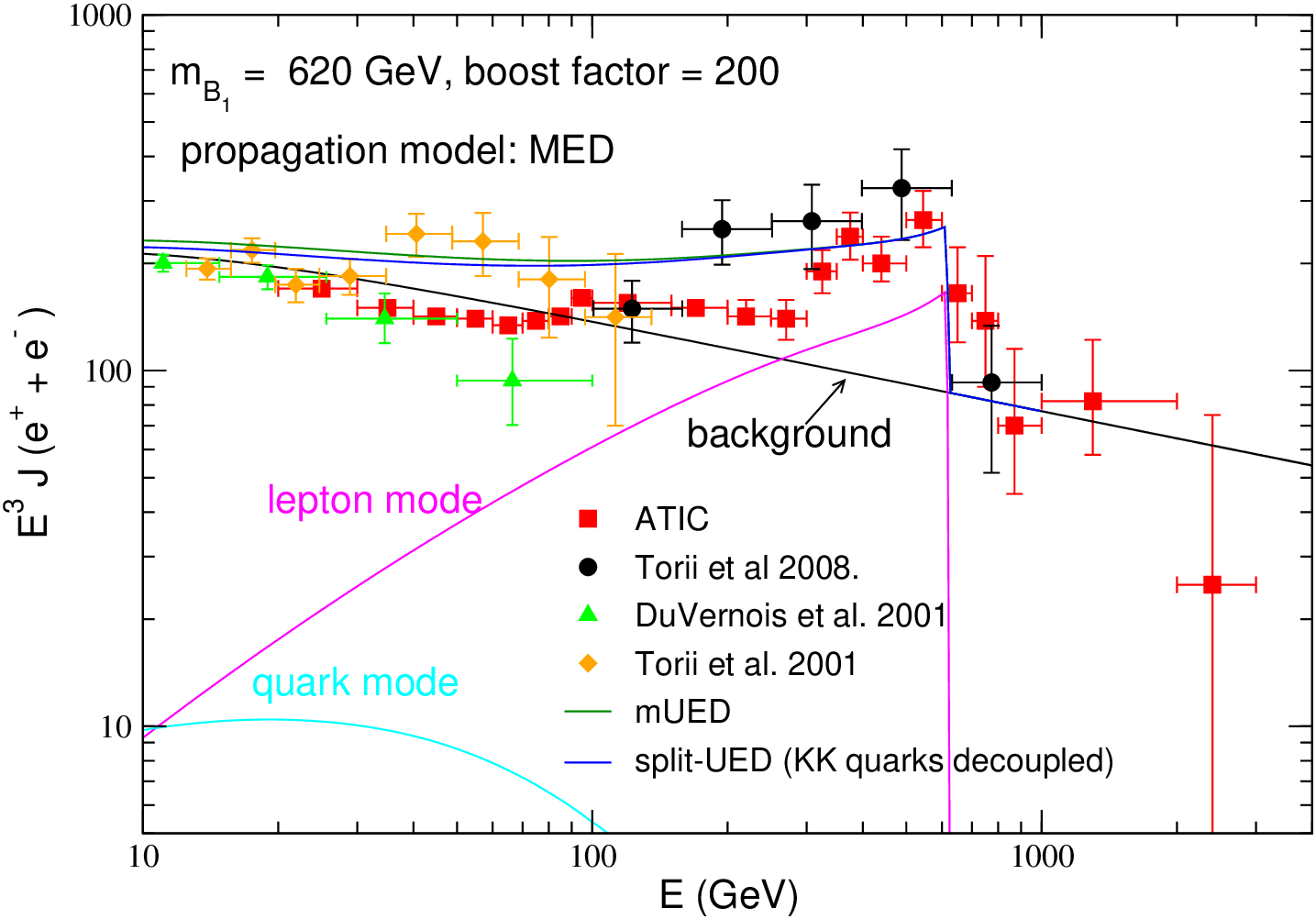}
\caption{ The flux of electron plus positron from LKP dark matter annihilation using MED diffusion model, compared with experimental data \cite{DuVernois:2001bb,Torii:2001aw,PPB-BETS,ATIC}. Note that only lepton mode is inclulded in the split-UED, since all the KK quarks are decoupled.}
\label{fig:atic1}
\end{figure}
respectively, where $k$ is a free parameter which is used to fit the data when
no primary source of $e^{+}$ flux exists \cite{Baltz:1998xv,Baltz:2001ir}. For our numerical simulation, we take the mass of dark matter, $m_{B_1}$ to be $620$ GeV. In Fig.\ref{fig:atic1}, we show the total flux of electron and positron with the background from Eq.(\ref{eq:bg_e}) and recent experimental data \cite{DuVernois:2001bb,Torii:2001aw,PPB-BETS,ATIC}. In order to have a better fit for the data, a boost factor of 200 is needed, which is consistent with the value (200) chosen in Ref.\cite{ATIC}. Boost factor is known to have origins such as local clumps in dark matter profile \cite{clump1} (see also \cite{clump2}), Sommerfeld enhancement effect by a  long range attractive force \cite{Sommerfeld1,Sommerfeld2,Sommerfeld3,Sommerfeld4,Sommerfeld5}, the Breit-Wigner type resonance effect \cite{Breit-Wigner} and possibly many more origins. In our case, without assuming a new attractive force or further tuning of mass spectra to get the large resonance, we tend to assume that boost factor mainly comes from clumps. We also show in Fig.\ref{fig:atic1} the contributions of quark mode and leptonic mode from the dark matter annihilation. The $e^{\pm}$ from quark mode  are much softer than that in the leptonic mode. This is because that the main source of $e^{\pm}$ in quark mode is the hadron cascade decay while in leptonic mode there exists a direct production of $B_1 B_1 \to e^+ e^-$, and $e^{\pm}$ from $\mu^{\pm}$ and $\tau^{\pm}$ decay are harder. We can easily see that the peak is mainly contributed by the leptonic mode and sharp drop-off is due to the direct production of $e^{\pm}$, therefore, the predictions of mUED and split-UED are quite similar to each other in the peak region since the difference between these two models is just in quark sector. We also compare different diffusion models in Fig.\ref{fig:atic2} for split-UED case. MED and M1 models produce more soft electrons and positrons than the M2 model does, thus the distributions of the former two models are flater. For energetic positrons and electrons, all of three models are quite the same with each other. 
\begin{figure}[t]
\includegraphics[width=0.75 \textwidth]{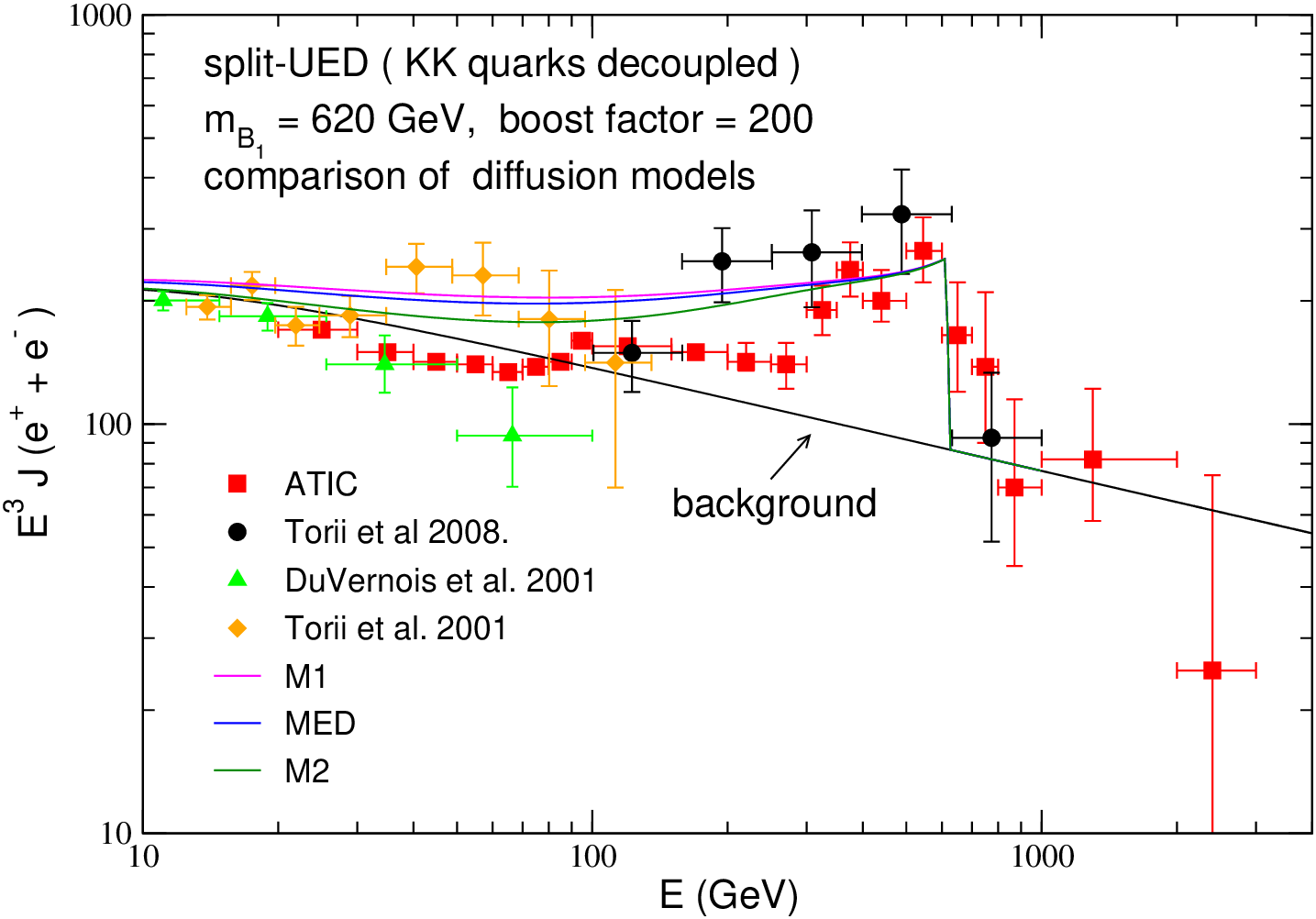}
\caption{The same as Fig.\ref{fig:atic1} but only for split-UED, and different diffusion models, M1, MED and M2, are shown together.}
\label{fig:atic2}
\end{figure}

\begin{figure}[t]
\includegraphics[width=0.75 \textwidth]{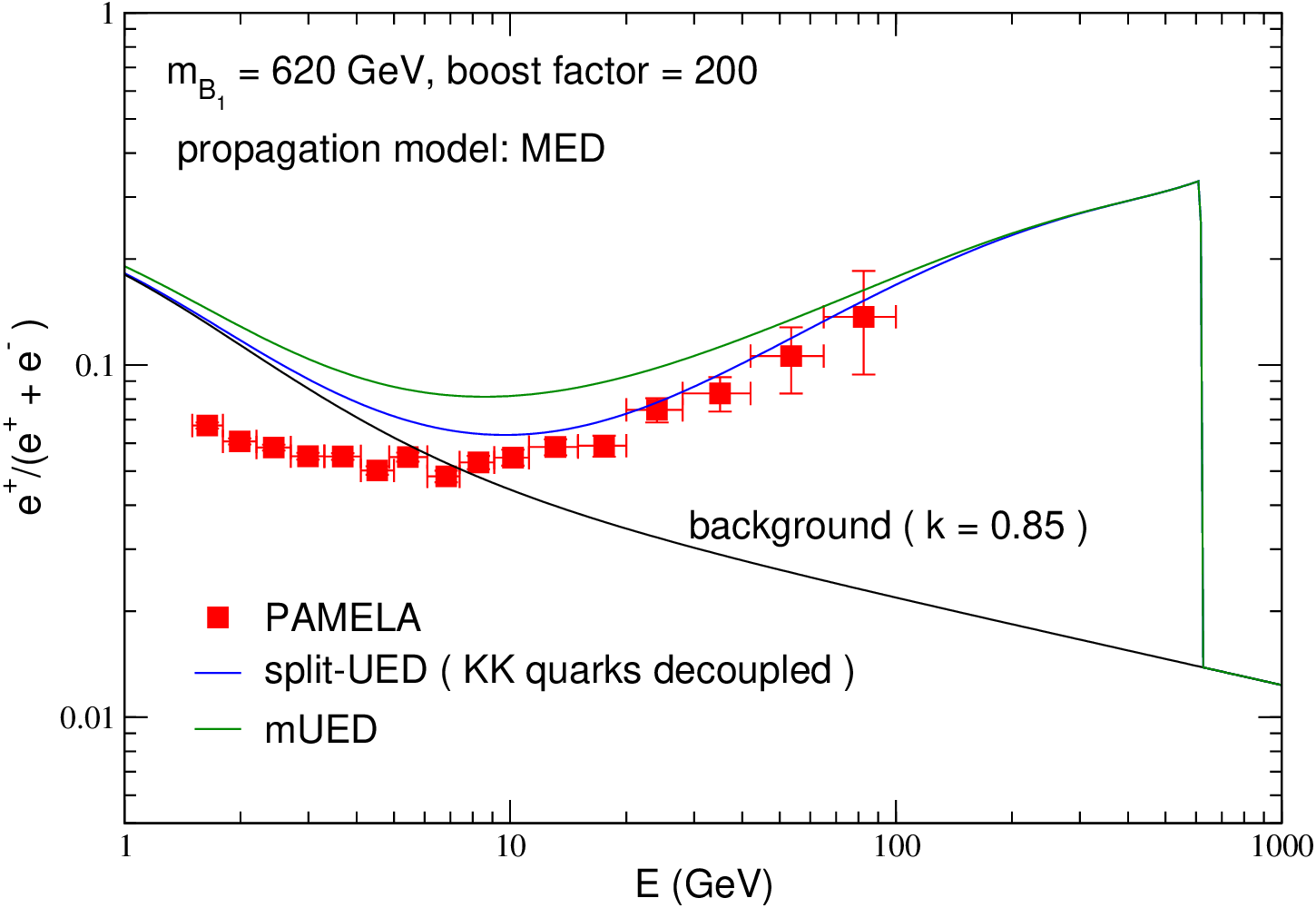}
\caption{The positron flux fraction predicted from the LKP dark matter annihilation using MED diffusion model, compared with PAMELA data \cite{PAMELA}.}
\label{fig:pamela1}
\end{figure}

\begin{figure}[]
\includegraphics[width=0.75\textwidth, angle=0]{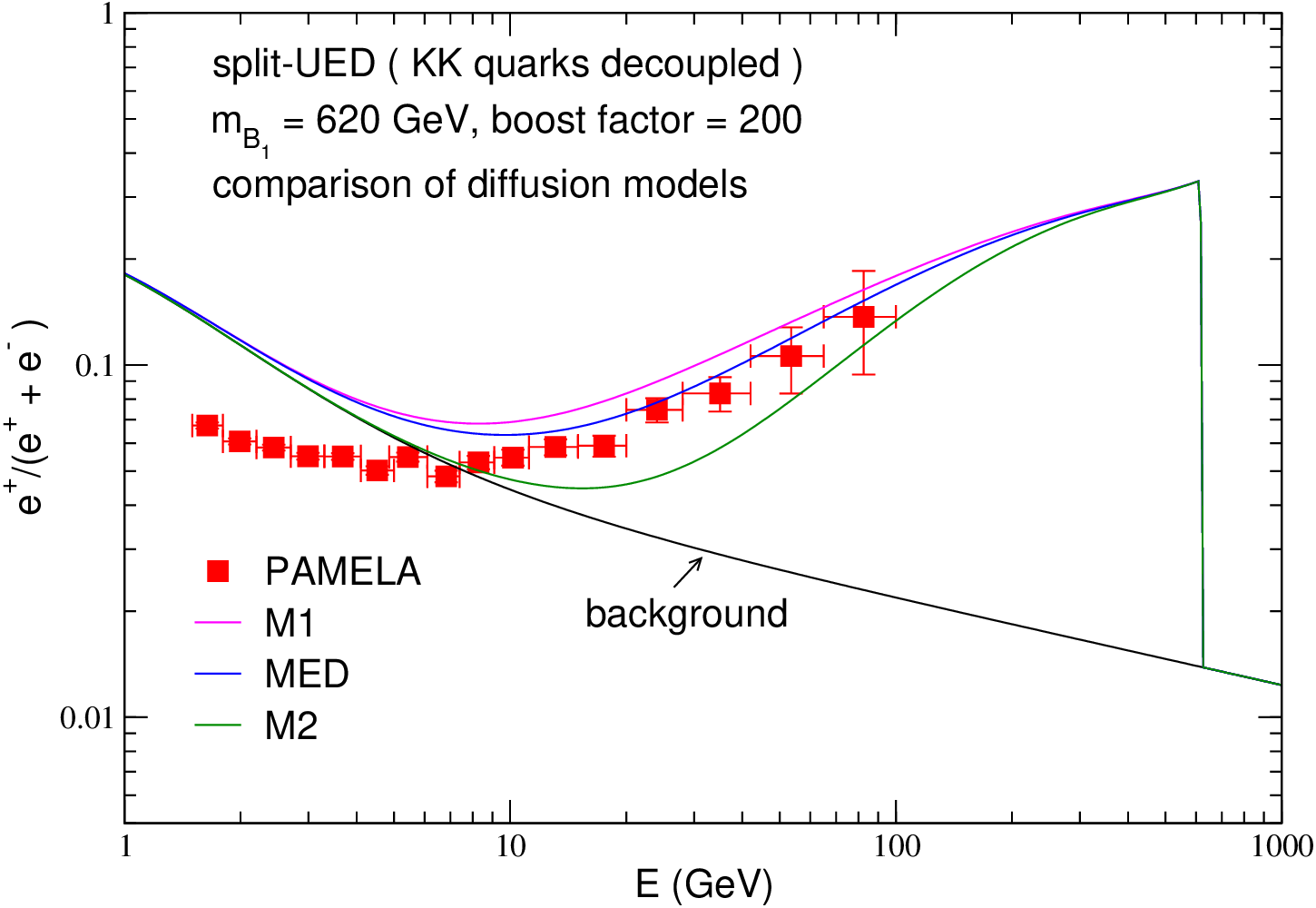}
\caption{Same as Fig.\ref{fig:pamela1} but only for split-UED, and different diffusion models, M1, MED and M2, are shown together.}
\label{fig:pamela2} 
\end{figure}

For positron fraction, we show in  Fig.\ref{fig:pamela1} the predictions of mUED and split-UED, using MED diffusion model,  with PAMELA data \cite{PAMELA}. Since there are more soft $e^+$ in mUED from the quark modes, the distribution is falter compared to that of split-UED.  If we adopt M2 and M1 models, as shown in  Fig.\ref{fig:pamela2}, the fittings seem worse compared the MED case for split-UED especially in the energy region around $10$ GeV. However, we have decoupled all of the higher KK quarks in this plot, i.e. hadronic final states are turned off in the dark matter annihilation. With finite mass of a first  KK quark, more soft positrons will be produced, therefore, the M2 model can also be consistent with the data. For M1 model, we expect that by tuning the parameters, e.g. boost factor or the normalization of the background, it would be possible to explain the PAMELA data as well. So, we conclude here that the dark matter in both mUED and spilt-UED can be the equivalently good  source of the excesses in PAMELA and ATIC/PPB-BETS data.      


\subsection{Antiproton}
\label{subsec:ap}
\begin{figure}[t]
\includegraphics[width=0.75 \textwidth, angle=0]{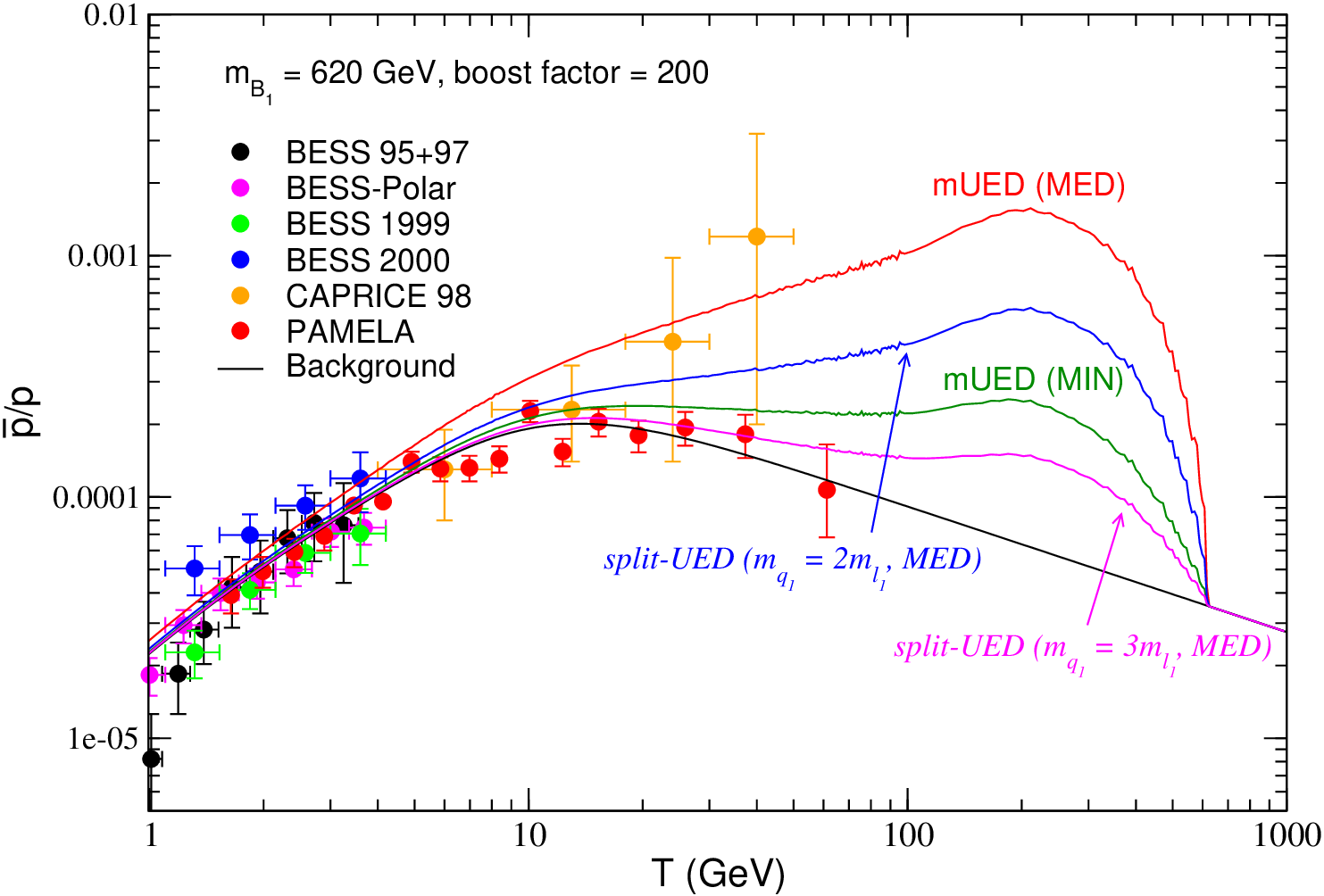}
\caption{ Antiproton to proton ratio predicted in mUED and split-UED models, together with experimental data. In mUED, both MED (red line) and MIN (green line) diffusion models are shown. For split-UED, the cases of $m_{q_1} = 2 m_{l_1}$ (blue line) and  $m_{q_1} = 3 m_{l_1}$ (magenta line) are presented, using MED diffusion model.}
\label{fig:ratio}
\end{figure}

The propagation of antiprotons through the Galaxy is similar to that of positrons.  The diffusion equation of antiprotons can be described as
\be
K_p\bigtriangledown^2 f_{\bar{p}}(T,\vec{r})-\frac{\partial}{\partial z}\left[( sign \, of \,z)V_c(z)f_{\bar{p}}(T,\vec{r})\right]-2h\delta(z)\Gamma_{ann} f_{\bar{p}}(T,\vec{r})+Q(T,\vec{r})=0,\,
\label{eq:diff_ap}
\ee
where $f_{\bar{p}}(T,\vec{r})$ is the number density of antiproton per unit energy, $T$ is the kinetic energy of antiproton and $K_p = K_{0p} \beta (p/\rm{GeV})^\delta$ is the diffusion parameter. $V_c(z)$ in the second term of Eq.(\ref{eq:diff_ap})  is related to the convective wind that tends to push antiprotons away from the Galactic plane, and is assumed to be a constant . Again, like the case in positron, the values of $K_{0p}$, $\delta$ and $V_c$ for different diffusion models are set to agree with the observed cosmic-ray data, especailly the B/C, and their values are listed in table \ref{tab:model_ap} \cite{Delahaye:2007fr}. The third term of Eq.(\ref{eq:diff_ap}) represents the annihilation of $\bar{p}$ with the interstellar proton in the Galactic plane, $h$ is the half-height of plane which is set to be $0.1$ kpc in our calculations. The antiproton annihilation rate $\Gamma_{ann}$ is given as \cite{Donato:2001ms}

\be
\Gamma_{ann}=(n_H + 4^{2/3}n_{He})\sigma^{ann}_{\bar{p}p}(T)v_{\bar{p}},
\label{eq:anni_ap}
\ee
where $n_{H}\sim 1/\rm{cm}^3$ and $n_{He}\sim 0.07 n_{H}$ are the number densities of hydrogen and helium, respectively; $\sigma^{ann}_{\bar{p}p}(T)$ is the annihilation cross section and $v_{\bar{p}}$ is the velocity of antiproton. The form of $\sigma^{ann}_{\bar{p}p}(T)$ is given by \cite{Tan:1983de}
\be
\sigma^{ann}_{\bar{p}p}(T)=\left\{ \begin{array}{cc}
661(1+0.0115T^{-0.774}-0.948T^{0.0151})\, \rm{mbarn}& ,\rm{for}\,T<15.5\,{\rm {GeV}};\\
36T^{-0.5}\,\rm{mbarn}\qquad\qquad\qquad\qquad\qquad\qquad\quad & ,\rm{for}\,T\geq15.5\,{\rm {GeV}}.\end{array}\right.
\ee
The solution of the interstellar flux of antiproton in the vicinity of solar system is \cite{Cirelli:2008id}
\bea
\Phi^{\rm{IS}}(T) & =&\frac{v_{\bar{p}}}{4\pi}f_{\bar{p}}(T,\vec{r_{\odot}})\\\nonumber
                  &=&\frac{v_{\bar{p}}}{4\pi}\frac{1}{2m_{B_1}^2}\sum_i<\sigma v>_i G(T)\left(\frac{dN_{\bar{p}}}{dT}\right)_i,
\eea
where the $dN_{\bar{p}}/dT$ is the spectrum of antiproton which we simulate with PYTHIA, and the index $i$ runs over all quark final states in the $B_1$ dark matter annihilation processes, and 
\be
G(T) = \sum_{n,m=1}^{\infty}J_0\left(\zeta_n \frac{r_{\odot}}{R} \right)\exp \left[-\frac{V_cL}{2K_p(T)} \right] \frac{y_n(L)}{A_n \sinh(S_n L/2)},
\ee
with
\be
y_n (L) = \frac{4}{J_1^2(\zeta_n)R^2}\int_0^R dr r J_0(\zeta_n r/R) \int_0^L dz \exp\left[ \frac{V_c (L-z)}{2K_p(T)} \right] \sinh(S_n(L-z)/2) \rho(\vec{r})^2,
\ee
and
\bea
A_n &=& 2h\Gamma_{ann} +V_c +K_p(T)S_n \coth(S_n L/2), \\
S_n &=& \sqrt{\frac{V_c^2}{K^2_p(T)}+\frac{4\zeta_n^2}{R^2}}.
\eea
However, we have to take solar modulation into account to estimate the flux of antiproton obtained at the the Earth, which is  important for the low energy antiproton. The flux of the antiprotons at the top of Earth's  atmosphere can, therefore, be written as \cite{ap:solar1,ap:solar2}
\be
\Phi_{\bar{p}}^{\rm{TOA}}(T_{\rm{TOA}})= \frac{2m_p T_{\rm{TOA}}+T_{\rm{TOA}}^2}{2m_p T_{\rm{IS}}+T_{\rm{IS}}^2}\Phi_{\bar{p}}^{\rm{IS}}(T_{\rm{IS}}),
\ee
 where $T_{\rm{TOA}}=T_{\rm{IS}}-\phi_F$ with $\phi_F$ being the solar modular parameter which we take to be $500$ MV in our calculation.
\begin{table}[t]
\begin{tabular}{|c|c|c|c|c|}
\hline 
Model&
$\delta$&
$K_{0p} (\rm{kpc^2/Myr})$&
$L (\rm{kpc})$&
$V_c (\rm{km/s})$\tabularnewline
\hline
\hline 
MIN&
$0.85$&
$0.0016$&
$1$&
$13.5$\tabularnewline
\hline
MED&
$0.70$&
$0.0112$&
$4$&
$12$\tabularnewline
\hline
MAX&
$0.46$&
$0.0765$&
$15$&
$5$\tabularnewline
\hline
\end{tabular}
\caption{The parameters in the diffusion models which are comparible with B/C and generate minimum (MIN), median (MED) and maximal (MAX) antiproton flux.  }
\label{tab:model_ap}
\end{table}
In order to compare with the observed data of antiproton to proton ratio, we need to include the astrophysical background. Here we adopt simple fittings for background antiprotons $\Phi_{\bar{p}}^{bg} $ and protons $\Phi_p^{bg}$ provided in Ref.\cite{Nezri:2009jd}
,
\bea
\Phi_{\bar{p}}^{bg} &=& \frac{0.9t^{-0.9}}{14+30t^{-1.85}+0.08t^{2.3}}\,[\rm{GeV}^{-1}\rm{m}^{-2}\rm{s}^{-1}\rm{sr}^{-1}],\\
\Phi_p^{bg} &=& 10^4\frac{0.9t^{-1}}{8+1.1t^{-1.85}+0.8t^{1.68}}\,[\rm{GeV}^{-1}\rm{m}^{-2}\rm{s}^{-1}\rm{sr}^{-1}].
\label{eq:ap_bg}
\eea
The predicted ratios of antiproton to proton, $\rpbp$, for mUED and split-UED are shown in Fig.\,\ref{fig:ratio} with experimental data \cite{Adriani:2008zq,Orito:1999re,Boezio:2001ac,Asaoka:2001fv,Abe:2008sh}. Here we take a common boost factor ($B_F=200$) which we have obtained from ATIC/PPB-BETS fit. Since the hard electronic signals are mainly from local sources within a few kpcs but antiprotons can come from distant sources, one may take different values of boost factor for positron and antiproton in one's calculation. Here, partly for predictability and simplicity and partly in order to follow common assumption that we live at a typical place  in our galaxy, we take a common boost factor for all particles in our calculations. The mUED agrees well in the low energy region $E\lesssim10$ GeV, but starts deviating significantly from the recent PAMELA data when energy of antiproton becomes higher, if MED diffusion model is adopted. Therefore, mUED seems to be disfavored by PAMELA data in the high energy region, and it agrees with PAMELA observations only when MIN diffusion model is chosen. However, one can easily see that how the situation can be improved in the split-UED case. By enhancing the mass of $q_1$ to be twice of the $l_1$, the prediction of $\rpbp$ is reduced by more than a factor of two, and will be much suppressed if $q_1$ becomes heavier, as shown by the blue and magenta lines in Fig.\ref{fig:ratio}. We also note that if we adopt MIN model, the split-UED case agrees with PAMELA very well even for $m_{q_1}=2m_{l_1}$. The lower bound of the mass of $q_1$  in split-UED can be estimated by the upcoming data of PAMELA in the higher energy region, since a small bump is predicted at $E\approx200$ GeV, and the height of deviation from background is controlled by the mass of $q_1$.

\subsection{Gamma-ray}
\label{subsec:gamma}
For diffusive gamma-ray, there are galactic and extragalactic contributions from the annihilation
of LKP dark matter $B_1$. The flux of the gamma-ray
from the extragalactic origin is estimated as \cite{Bergstrom:2001jj}

\begin{equation}
\left[E^{2}\frac{dJ_{\gamma}}{dE}\right]_{eg}=\frac{B_F E^{2}c\Omega_{B_1}^2\rho_{c}^2}{4\pi m_{B_1}^2H_{0}\Omega_{M}^{1/2}}\sum_i <\sigma v>_i\int_{0}^{z_{{\rm up}}}dy\left(\frac{dN_{\gamma}}{d((1+z)E)}\right)_i\frac{(1+z)^{-3/2}e^{-z/z_{\rm{max}}}}{\sqrt{1+\frac{\Omega_{\Lambda}}{\Omega_{M}}(1+z)^{-3}}},
\label{eq:gamma_eg}
\end{equation}
where $\Omega_{B_1}$, $\Omega_{M}$ and
$\Omega_{\Lambda}$ are the density parameters of $B_1$, matter (including
both baryons and dark matter) and the cosmological constant, respectively;
$\rho_{c}$ is the critical density; $H_{0}$ is the Hubble parameter at the present time; $z$ is the redshift, and $z_{up}=m_{B_1}/E-1$ since the maximum energy of photon in the rest frame of annihilation is $E = m_{B_1}$; the term $e^{-z/z_{\rm{max}}}$ with $z_{\rm{max}}(E)\sim3.3(E/10\rm{GeV})^{-0.8}$ takes the optical depth into account \cite{Bergstrom:2001jj}. For the numerical results,
we use\ \cite{Komatsu:2008hk}\begin{equation}
\Omega_{B_1}h^{2}=0.1099,\quad\Omega_{M}h^{2}=0.1326,\quad\Omega_{\Lambda}=0.742,\quad\rho_{c}=1.0537\times10^{-5}{\rm GeV/cm}^{3}.\,\label{eq:input}\end{equation}

On the other hand, the gamma-ray flux from the annihilation of $B_1$ in
the Milky Way halo is 
\begin{equation}
\left[E^{2}\frac{dJ_{\gamma}}{dE}\right]_{halo}=\frac{E^{2}}{4\pi}\frac{1}{2m_{B_1}^2}\frac{dN_{\gamma}}{dE}\left\langle \int_{los}\rho^2(\vec{\ell})d\vec{\ell}\right\rangle ,\label{eq:gamma_halo}\end{equation}
 where $\rho_{halo}$ is the density profile of dark matter in the
Milky Way, $\left\langle \int_{los}\rho^2(\vec{\ell})d\vec{\ell}\right\rangle $
is the average of the integration along the line of sight (los). 

\begin{figure}[t]
\includegraphics[width=0.75 \textwidth]{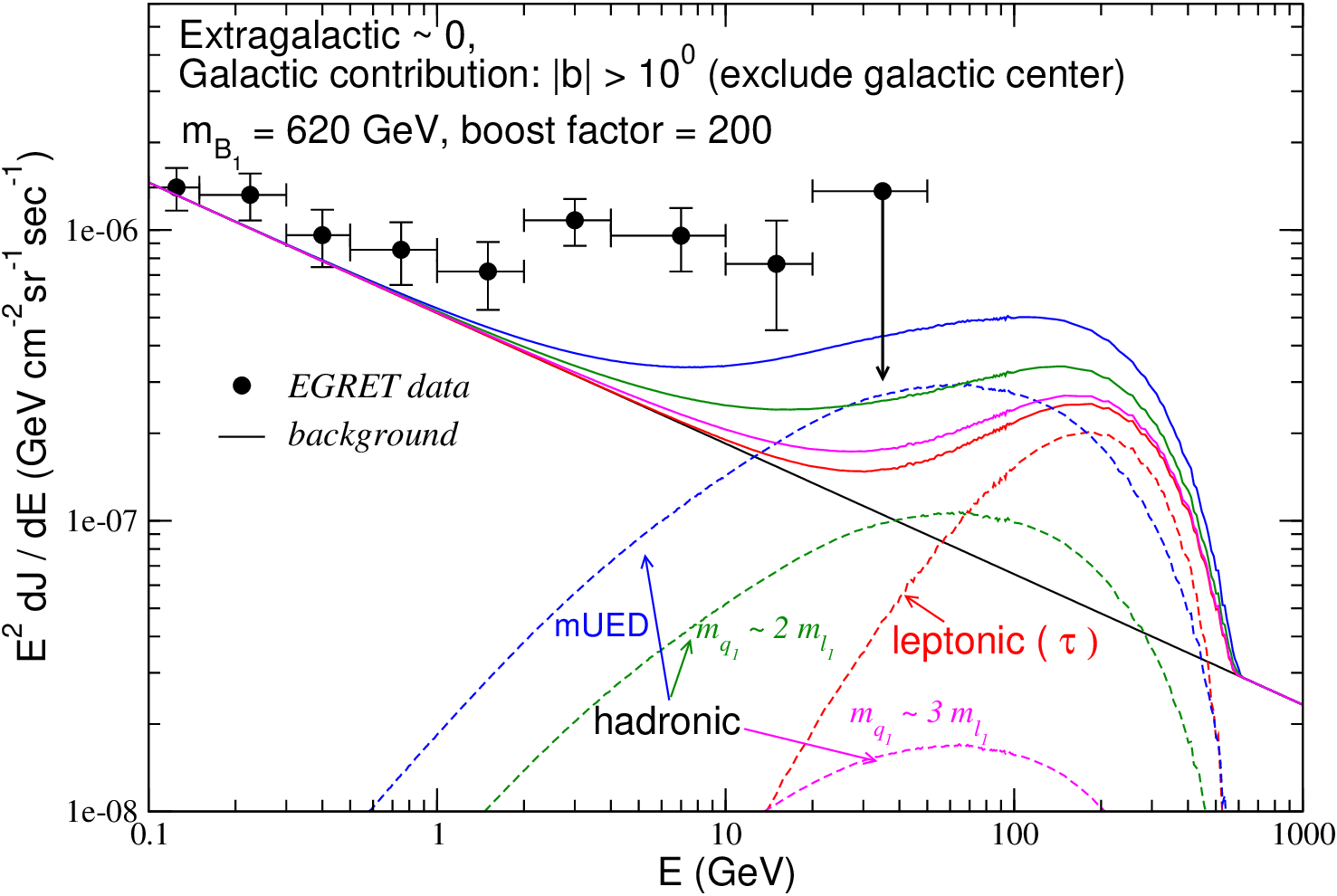}
\caption{ The diffusive gamma-ray predicted in mUED and split-UED models, together with EGRET data. The solid lines are for signals with background, blue line is for mUED; green, magenta and red lines are for $m_{q_1}=2m_{l_1}$, $m_{q_1}=3m_{l_1}$ and $m_{q_1}\to \infty$, respectively. Signals of dark matter annihilation are shown separately in dashed lies for leptonic final state (red), hadronic final states in mUED (blue) and in split-UED with $m_{q_1}=2m_{l_1}$ (green) and $m_{q_1}=3m_{l_1}$ (magenta).}
\label{fig:gamma}
\end{figure}

Since there are still big uncertainties and ambiguities for modeling the Galaxy center, we integrate over the whole sky except for the zone of the
Galactic plane (i.e. exclude the region with the galactic latitudes $|b|<10^{\circ}$).
For the background, we use a power-law form adopted in Ref.\ \cite{Ishiwata:2008cu}
\begin{equation}
\left[E^{2}\frac{dJ_{\gamma}}{dE}\right]_{bg}\simeq5.18\times10^{-7}E^{-0.499}\quad{\rm GeVcm}^{-2}{\rm sr}^{-1}{\rm sec}^{-1},\label{eq:gamma_bg}\end{equation}
 where $E$ is in units of GeV. 

The total flux of diffusive gamma-ray is shown in Fig.\ref{fig:gamma}, and the signals from the hadronic and leptonic final states in $B_1 B_1$ annihilation are presented as well. We should emphasize that we do not try to fit EGRET data ~\cite{Sreekumar:1997un,Strong:2004ry} in the plot, since  Fermi ~\cite{FGST} is expected to have  more precise results soon, however, we still show EGRET data for reference. We notice that the extragalactic contribution from cosmological distance is very small even we enhance it by a factor of $10^6$ to estimate the effect of subhalo \cite{Bergstrom:2001jj}. The gamma-ray in the decay of $\pi^0$ from hadron cascade decay is softer than that from the decay of $\tau$, therefore the distribution become flater when more quarks are produced in $B_1$ annihilation. By suppressing the fraction of the quark mode in the final state, i.e. comparing mUED (blue lines) and split-UED (green, magenta and red lines), the starting point of deviation from the background will shift to high energy \footnote{The preliminary result of Fermi \cite{fermiTalk} shows the data is consistent with the estimated background up to $E\sim 10$ GeV, so by enhancing the mass of $q_1$ in split-UED seems to be prefered, if the preliminary result is thoughtful.} and the predicted peak at energy about 200 GeV will become smaller. The bump at the high energy is a generic prediction for split-UED even all of the higher order KK quarks are decoupled, because  gamma-rays still come from the $\tau$ decay, and it can be further examined by the Fermi experiment in the near future.

\section{Collider signature of split-UED}
\label{sec:collider}

Having light colored particles below a few TeV in split-UED, the LHC can produce lots of KK quarks and gluons via QCD interactions. 
As it is already shown in Sec.~\ref{sec:model}, one of the 
features of our model is that KK quarks are split from the other particles, therefore, will lead to collider phenomenology quite different from the mUED models. For example, the 1st KK quark, where we take $m_{q_1}\approx2m_{B_1}$ through calculations, mainly decays to the 1st KK gluon $g_1$ and a SM quark, generating at least one high $p_T$ jet due to the big mass gap between $q_1$ and $g_1$. Another one is the existance of tree-level KK number non-conserved interactions between KK-even gauge bosons and the SM fermions, as a result, the production cross sections of 2nd KK gauge bosons can be substantial. It has been shown that the signals of a 2nd KK gluon production can be dominant over the SM background in dijet events after imposing a certain set of cuts at the LHC \cite{split-UED}. In this section, we will focus on the scenarios of 1st KK quarks and gluon, which can only be produced in pairs.  

 \begin{figure}[t]
\includegraphics[width=0.45 \textwidth, angle=90]{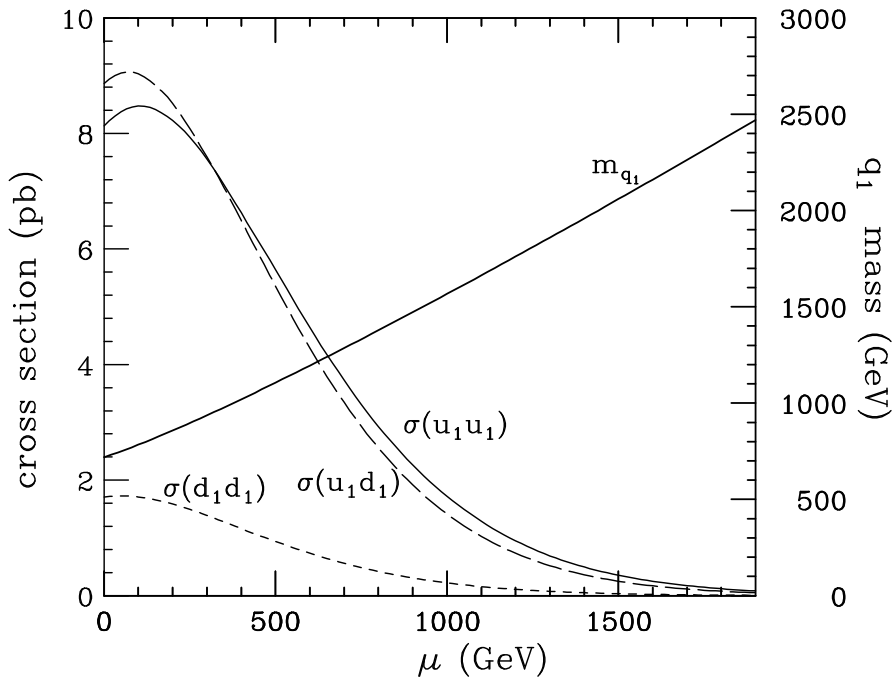}
 \caption{The cross sections and the mass of $u_{L1}$ as functions of the bulk mass $\mu$ with 
 fixing $1/R=620$ GeV.
Here, $u_1$($d_1$) includes both $u_{L1}$ and $u_{R1}$ ($d_{L1}$ and $d_{R1}$) contributions. }\label{col1}
\end{figure}

The production cross sections involving $q_1$ at the LHC 
 depend on the bulk mass parameter $\mu$. 
Among the production cross sections of $q_1 q_1^{(\prime)}$, $q_1 \bar{q}_1^{(\prime)}$ and 
$\bar{q}_1\bar{q}_1^{(\prime)}$,  the $\sigma(q_1 q_1) $ is dominant one 
as the valence  $u$ and $d$ partons can contribute 
to the initial states. 
These productions proceed mainly through the $t$-channel diagrams with the 1st KK gauge bosons exchanged. The most relevant couplings  for these productions are $q_1$-$q$-$V_1$, where $V_1 = g_1$, $B_1$, the 1st KK $W-$boson $W_1$  and the 1st KK $Z-$boson $Z_1$. 
The cross section $\sigma(u_1 u_1)$, $\sigma(u_1 d_1)$ 
and $\sigma(d_1d_1)$ are plotted in Fig.~\ref{col1} with varying $\mu$, and we neglect the $q_1\bar{q}_1^{(\prime)}$ productions, since they are insignificant.  
For small $\mu$, the production cross sections increase with increasing $\mu$
due to the enhancement from the couplings, as shown in Fig.\ref{fig3}, since the cross section is  proportional to  $ ({\cal G}_{110})^4$.  
For large $\mu$,  on the other hand, those cross sections decrease very quickly
with increasing $\mu$ (increasing $m_{q_1}$) as expected. 
For the $g_1q_1$ and $g_1 g_1$ productions, the cross sections, which are not shown, are at the same order as that of $q_1q_1$. Here all cross sections are calculated using {\tt CalcHEP} \cite{Pukhov:2004ca} by modifying the mUED model file in Ref. \cite{mUEDcalc}.

The signature of split-UED $q_1q_1$   productions followed by 
$q_1 \rightarrow g_1 q$ $\rightarrow B_1  qX$ is two high $p_T$  jets plus soft jets/leptons 
and large missing momentum.  
The kinematics and the decay branching ratios 
are  very similar to those of the corresponding supersymmetric model (SUSY) processes, namely  
the $\tilde{q}\tilde{q}$ productions followed by   $\tilde{q}\rightarrow \tilde{g} q$ $\rightarrow \tilde{\chi}_1^0 qX$
with squark mass $m_{\tilde{q}}=m_{q_1}$, gluino mass $m_{\tilde{g}} =m_{g_1}$ and the lightest neutralino mass   $m_{\tilde{\chi}^0_1}=m_{B_1}$. Therefore, we mimic split-UED collider signatures of $q_1$ pair productions at the LHC by scaling the total cross sections of SUSY signatures with the model point whose mass spectrum is similar to that of the split-UED. Furthermore, because collider signatures of $q_1 q_1$ are the same as that of $\tilde{q}\tilde{q}$, they share the same SM background. Since the SM background for squark and gluino productions  at the LHC is  very well studied \cite{ATLASEP}, we apply the same cuts in our calculation. The model points we study are listed in Table \ref{colt1}, in which we use ISAJET \cite{Paige:2003mg} for SUSY, and we take $1/R$ and $\mu$ to be $620$ GeV and $700$ GeV, respectively, for split-UED so that $m_{q_1} \approx 2m_{B_1}$. 
The events are generated by using {\tt HERWIG} \cite{Corcella:2002jc} for SUSY processes, 
then we regard them as the split-UED events,
and detector simulations are 
carried out by {\tt AcerDET} \cite{Richter-Was:2002ch}. Table \ref{colt2} shows the acceptances of the events after imposing the following basic cuts:
\be
M_{\rm eff}> 500 ~{\rm GeV},\,\, E_{\rm Tmiss}> \max (100~{\rm GeV}, 0.2M_{\rm eff} ),\,\,n_{100}\ge 1,\,\,n_{50}\ge 4,
\label{eq:cuts}
\ee
$M_{\rm eff}$ is the scalar sum of the $p_T$ of the
 first four leading jets; $E_{\rm Tmiss}$ is the transverse missing momentum; $n_{100}$ ($n_{50}$) 
 is the number of jets with $p_T> 100$ $(50)$~GeV. Forthermore, we also use harder cuts on $M_{\rm eff}$ with $M_{\rm eff} > 1,\,1.5 \rm{TeV}$.

\begin{table}
\begin{center}
\begin{tabular}{|c|c||c|c|}
\hline 
 split-UED &   mass        &SUSY &mass\cr
\hline
\hline &&&\\[-12pt]
  $q_{L1}$&1347~GeV  & $\tilde{u}_L$ ,$\tilde{d}_L$& 1355, 1358 ~GeV  \cr
  $u_{R1}$&1322~GeV  & $\tilde{u}_R$ & 1304~GeV \cr
  $ d_{R1} $& 1318~GeV  & $\tilde{d}_R$  &  1263~GeV \cr
  $ g_1$     &\  794~GeV  & $\tilde{g}$      &\ 799~GeV \cr
  $B_1$      &\  621~GeV & $\tilde{\chi}^0_1$&\  622~GeV \cr
\hline 
  \end{tabular}
\end{center}
\caption{Mass spectrum of split-UED at $1/R=620$~GeV and $ \mu=700$~GeV. The 
corresponding SUSY point generated using  ISAJET is also listed. }\label{colt1}
\end{table}

 \begin{table}
\begin{center}
\begin{tabular} {|r|r||r|r|}
\hline
         & after standard cut & $M_{\rm eff}>1$\,TeV &
          $M_{\rm eff}>1.5$\,TeV\cr
           \hline 
           \hline 
$q_1 q_1$ & 0.40  & 0.37  & 0.21\cr
$q_1 g_1$  & 0.30 & 0.18  &0.049\cr
$g_1 g_1$ & 0.18 & 0.04 & 0.007\cr
\hline 
\end{tabular}
\end{center}
\caption{The acceptances for the split-UED $q_1q_1$, $q_1g_1$ and $g_1g_1$ productions after imposing basic event selection cuts (\ref{eq:cuts}) and harder $M_{\rm eff}$ cuts for the model point in Table \ref{colt1}. Note that these numbers are obtained from SUSY events.}\label{colt2}
\end{table}

\begin{figure}[t]\begin{center}
 \includegraphics[width=0.32 \textwidth]{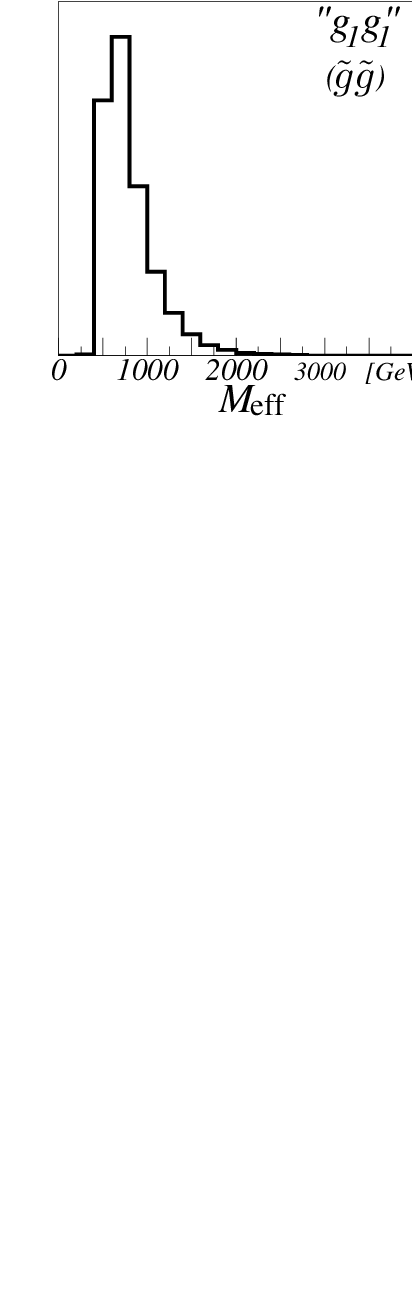} 
 \includegraphics[width=0.32 \textwidth]{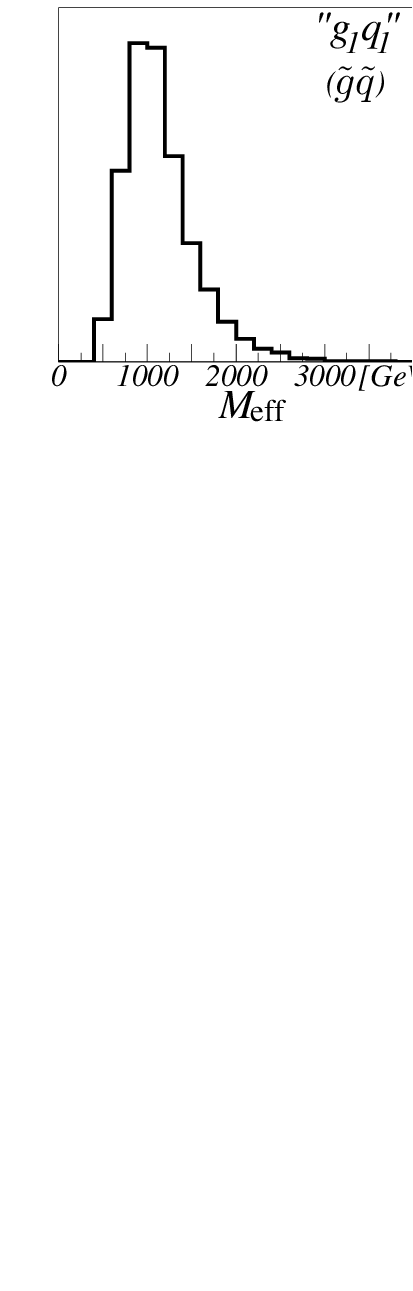} 
\includegraphics[width=0.32 \textwidth]{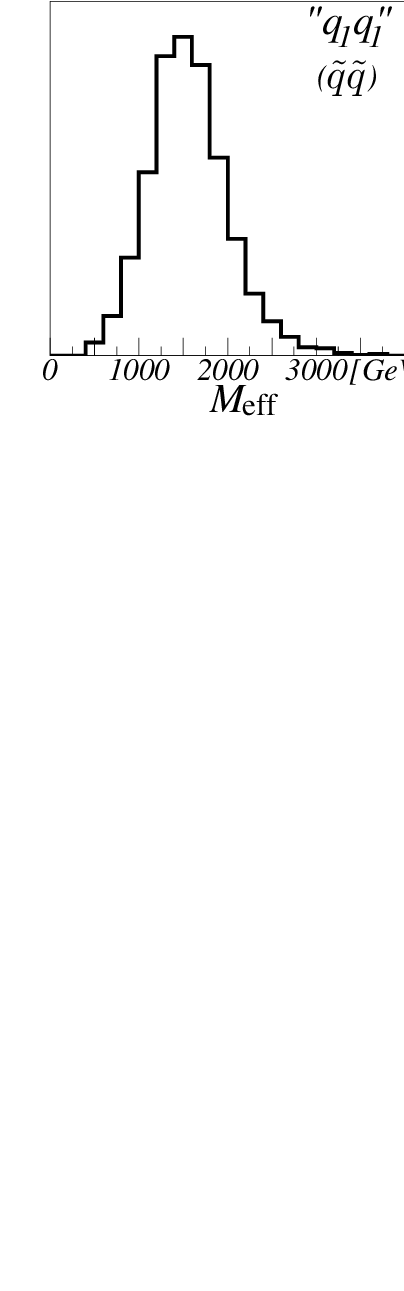}
 \end{center}
 \caption{$M_{\rm eff} $ distributions for ``$g_1g_1$", ``$q_1g_1$" and ``$q_1q_1$" from 
 left to right obtained by corresponding SUSY channels $\tilde{g}\tilde{g}$, $\tilde{q}\tilde{g}$ and $\tilde{q}\tilde{q}$. We neglect the y-axis here since we focus on the shapes, and we see that the distributions shift to higher energy region from left to right. }
 \label{col5}
 \end{figure}

 We can see that the acceptance  
 is about 40 \% for $q_1$ pair production after the basic cuts. 
 On the other hand, the efficiency to select the $g_1$ pair production is much smaller, especially  when a large $M_{\rm eff}$ is required. 
 For example, the number of events from the 1st KK gluon pair production reduces by factor of 1/30 after imposing the basic cuts with $M_{\rm eff}>1$ TeV.  
 This is because the probability of having high 
 $p_T$  jets and large $E_{\rm Tmiss}$  is very low for the $g_1$ production events at our point. Thus, we  expect $g_1g_1$ 
production is not promising due to the very small efficiency found in 
the simulation. 
The $g_1q_1$ production may be
  easier to detect compared to the $g_1g_1$ case with the help of the  high $p_T$ jet  from  $q_1\rightarrow q g_1$, 
  although the signal and the background separation would be  
  worse compared with $q_1q_1$ production.   
  Therefore we only consider  $q_1q_1$ productions in the following studies. For completeness, we show the  $M_{\rm  eff}$ distributions for $g_1g_1$, 
   $q_1g_1$ and $q_1q_1$  in Fig.~\ref{col5}. Again these results are obtained from SUSY events mimicking split-UED with the same kinematics.
   %

  The total  production cross section of $q_1q_1$, including $u_{L1}$,$ d_{L1}$, $u_{R1}$, 
 $d_{R1 }$, is 7.64~pb, and  we expect 7640 events for 1 fb$^{-1}$.
According to Table \ref{col5}, we expect $7640 \times 0.37= 2830$ events left 
under the standard cut with $M_{\rm eff} > 1$TeV.
We should also consider lepton veto before using the background studied in Ref.\cite{ATLASEP}.
Although it depends on the lepton branching ratio, we expect
more than a half number of events, i.e. 1400, remain with the lepton veto.
 The number of SM background events with the same cut for 1fb$^{-1}$ is 
less than 300
according to the $M_{\rm eff}$ distribution shown in Ref.~\cite{ATLASEP}.
Therefore the $q_1q_1$ signal distribution is well above the background distribution 
for $M_{\rm eff} > 1000$ GeV.

 \begin{figure}[t]
\begin{center}
 \includegraphics[width=0.65 \textwidth]{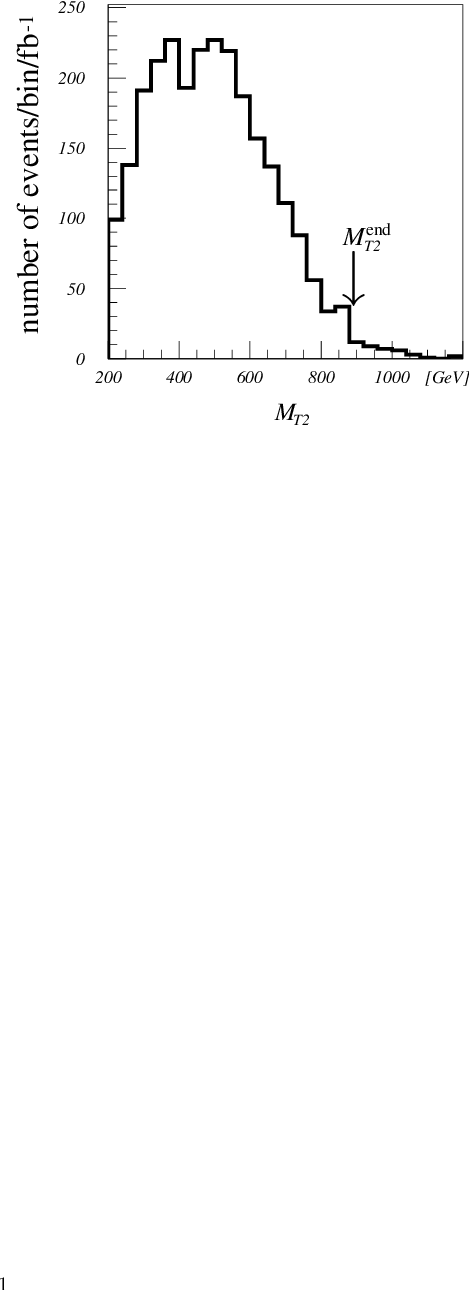}
 \end{center}
 \caption{The $M_{T2}$ distribution for $q_1 q_1$ production events under the standard cut with 
 $M_{\rm eff} > 1$ TeV for 1 fb$^{-1}$.
We can see that 
the end point of $M_{T2}$ distribution is given by
$m_{q_1}-\frac{m_{g_1}^2}{m_{q_1}}\simeq 880$~GeV. 
The bin size is 10 GeV/bin.
Note that we do not apply any lepton veto cut for this plot.
}\label{col4}
\end{figure}

Given the enough statistics, we now  discuss the possibility 
to determine  some of the split-UED  particle masses. 
 Fig.  \ref{col4} shows the $M_{T2}$ \cite{Barr:2003rg} distribution of the two highest 
$p_T$ jets  under the standard  cut  with $M_{\rm eff}> 1000$ GeV. The $M_{T2}$ is usually used to determine the masses of unknown particles which are pair-produced and decay identically. In our case, we have two $q_1$'s being produced, and both decay into $g_1 q$.
   Here the $M_{T2}$ is calculated 
  from the two highest $p_T $ jets,  $ j_1$ and $ 
  j_2$ and a missing transverse momentum  defined as  
$p_{T\rm miss }= - p_{T  j_1} -p_{T j_2}$ and the  formula is 
\begin{equation}
M_{T2}= \min_{p_{T{\rm miss}}=q_{T1}+ q_{T2}} \left[\max \{M_T(q_{T1}, p_{ j_1},m_{\rm trial}), M_T(q_{T2}, p_{ j_2},m_{\rm trial})\}\right],
\end{equation}
where $q_{T1}$ and $q_{T2}$ are two dummy parameters 
that make up $p_{T{\rm miss}}$,  and the minimization is taken for all possible sets of $q_{T1}$ and $q_{T2}$; $p_{j_1 (j_2)}$ is the momentum of $j_{1(2)}$; $m_{\rm trial}$ is the trial mass that represents the unknown mass of the daughter particle ($g_1$ in our process) and $M_T$ is the transverse mass. The $M_{T2}$ depends on the trial mass $m_{\rm trial}$, and we took $m_{\rm trial}=0$ in Fig.\ref{col4}.  
  For our simulation, the two highest $p_T$ jets comes from $q_1\rightarrow g_1 q$, 
and the end point of
  $M_{T2}$ should  be $m_{q_1}$ when the trial  particle mass $m_{\rm trial}$ is taken 
  as the true mass of $g_1$.  
  Another important fact is that when there are no initial state radiations, the formula  of 
  the $M_{T2}$ endpoint  for  $m_{\rm trial}=0$ has a general form given as 
  \begin{equation}
 M^{\rm end }_{T2}=m_A-\frac{m_X^2}{m_A},
 \end{equation}
 where $m_A$ is the mass of mother particle which is pair produced and $m_X$ is the mass of the unknown daughter particle, and in our case $m_A = m_{q_1}$ and $m_X=m_{g_1}$.
We can see that the $M_{T2}$ distribution in Fig. \ref{col4} has a clear  end point which 
  is consistent with the  predicted value $M_{T2}^{\rm end}\simeq$ 880~GeV for the model point in Table \ref{colt1} . Although we are unable to measure precise masses of $q_1$ and $g_1$, we still have a useful information on their combination. 

Finally, let us comment on the comparison with mSUGRA.
If we assume the signature comes from squark production, 
with a typical mass spectrum 
$m_{\tilde{q}} \sim $ 1 TeV 
and gives the similar $M_{T2}$ end point,
 the cross section is much smaller than that of the split-UED point. Thus, SUSY and split-UED should be distinguished  based on the event rates, although the kinematical nature of the signal is similar. 
We can get a rough understanding on the enhancement of the cross section in split-UED by comparing the helicity structure of squarks/gluinos and corresponding KK quarks/gluons \cite{Cheng:2002ab}. \footnote{See also \cite{Kong:2007uu} for similar phenomenon in Little Higgs model.}

\newpage
\section{Conclusion}
\label{sec:conclusion}

The split-UED model proposed in Ref. \cite{split-UED} is studied in detail. From the basic set-up of the model, we calculate all the masses (including one-loop radiative corrections to the masses) and relevant couplings for the first KK particles. Characteristic split spectrum of KK quark states is realized and enables us to control the hadronic branching fraction of the LKP annihilation by one parameter, bulk mass scale ($\mu$). In the non-relativistiv limit, such a branching fraction decreases very fast by increasing $\mu$.

In order to explain the anomalies from the cosmic ray, we fixed the size of extra dimension ($1/R \simeq 620$ GeV) from the peak position in flux of electrons plus positrons observed by ATIC and PPB-BETS, with boost factor $B=200$. Then all data from PAMELA, ATIC, and PPB-BETS can simultaneously be fitted very well. It is important to notice that such a dark matter mass in split-UED generically can explain the right amount of relic density of  dark matter ($\Omega \simeq 0.23$). To avoid all the strong constraints known up to date, we take $m_{q_1}\gtrsim2/R$ so that the hadronic branching fraction of LKP annihilation is significantly smaller than the leptonic one ($\sigma_{q\bar{q}}/\sigma_{l\bar{l}}<10\%$) and the cosmic antiproton flux is suppressed more than the safe rate. 
The flux of cosmic gamma-rays from pion decay is also highly suppressed and hardly detected  in low energy region ($E_\gamma\lesssim 20$ GeV). However future coverage of high energy domain in $O(100)$ GeV will make it possible to detect gamma-rays originated by tau lepton which is still sizable.   

At the LHC, the 1st KK colored particles (KK quarks and gluons) in split-UED are copiously produced and further decay into hard jets, soft jets/leptons and dark matter (missing energy).  Because of the large mass splitting between the 1st KK quark $q_1$ and 1st KK gluon $g_1$, we can separate out the production of $g_1 g_1$, $g_1 q_1$ and $q_1 q_1$ by using the $M_{\textrm{eff}}$ cuts.
We focus on the $q_1q_1$ production,in particular, which has  two hard jets so that we can distinguish the signals from the SM background. The split-UED $q_1$ pair signal is simulated by scaling the SUSY signatures with the same mass spectrum. After the appropriate cuts, our signal is well above the SM background. We calculate the $M_{T2}$ distribution of the $q_1$ pair signals, and its end point reflects the information for combination of $q_1$ and $g_1$ masses. Although we are unable to determine the mass of individual particle in split-UED, its predictions of the signatures of $q_1q_1$ production at the LHC, e.g. $M_{\rm eff}$ and $M_{T2}$, can be examined whether they are consistent with the cosmic-ray signatures in the near future, and vice versa.

\section{Acknowledgement}
C.R.C. thanks F. Takahashi for useful discussions. This work was supported by the World Premier International Research Center Initiative 
(WPI initiative) by MEXT, Japan.

\section*{Appendix A: KK decomposition in split-UED}

For all the gauge bosons, scalars, and leptons, their KK decompositions are the same as those in mUED. For bosons and fermions that are in the same chirality of the zero mode, which has $(+,+)$ boundary condition at $[-L, L]$, the KK decomposition is:
\bea
\label{eq:even}
\Phi_+(y) = \frac{1}{\sqrt{2L}}\phi_0(x) + \frac{1} {\sqrt{L}} \sum_{n^+, n^-} \sin\left[\frac{n^-  \pi}{2 L} y \right] \phi_{n^-}(x) + \cos \left[\frac{n^+ \pi} {2 L} y \right] \phi_{n^+}(x)  \ .
\eea
While for fermions that are in the opposite chirality of the zero mode, which has the $(-,-)$ boundary condition at $[-L, L]$, the KK decomposition is:
\bea
\label{eq:odd}
\Phi_-(y) =  \frac{1} {\sqrt{L}} \sum_{n^+, n^-} \cos \left[\frac{n^-  \pi}{2 L} y \right] \phi_{n^-}(x) + \sin \left[\frac{n^+ \pi} {2 L} y \right] \phi_{n^+}(x) \ .
\eea
The label $n^-$ or $n^+$ here stands for the $n$-th KK modes with the even/odd KK parity\footnote{In our convention, the KK number $n^- = 2n^- -1$ and $n^+ = 2 n^+$ for the $\{n^-, n^+ \}$ in Ref. 
\cite{Agashe:2007jb}.}. 

For the quark, we consider the case in which a SM quark is embeded into $\Psi_+= P_+ \Psi$ component of a 5D Dirac fermion $\Psi$, where $P_+ = (1+ \gamma_5)/2$ is the positive chirality projection operator. The other case could be considered quite similarly by replacing $\mu$ by $-\mu$. A convenient way of expressing the KK decomposition is in the form \cite{split-UED, Agashe:2007jb}:
\bea
\label{eq:quarkKK}
\Psi_+(x,y) &=& \sum_{n^+, n^-} g_{n^+}(|y|) \chi_{n^+}(x) + \epsilon(y) g_{n^-}(|y|) \chi_{n^-}(x), 
\nonumber \\
\Psi_-(x,y) &=& \sum_{n^+, n^-} \epsilon(y) f_{n^+}(|y|) \psi_{n^+}(x)+  f_{n^-}(|y|) \psi_{n^-}(x). \eea
The 5D profiles satisfy the following coupled, first-order equations of motion 
\begin{eqnarray}
\partial_y g_n + \mu g_n - m_n f_n&=& 0, \nonumber \\
\partial_y f_n - \mu f_n + m_n g_n&=& 0 ,
\end{eqnarray}
with each 5D profile $g_{n^+}$, $g_{n^-}$, $f_{n^+}$, $f_{n^-}$ satisfying the $(+,+)$, $(-,+)$, $(-,-)$, $(+,-)$ boundary condition at $y = 0$ and $L$, respectively. 
Once the solution for $y \subset [0,L]$ is obtained, the solution to the whole space is determined from Eq. (\ref{eq:quarkKK}) thanks to the symmetry. 

For the zero modes, $m_n=0$, the equations are separable and the solution is given as
\begin{eqnarray}
\label{eq:quarkzero}
g_{0^+}(y) = A_0 \exp (\int_0^y \lambda  \langle \Phi(s) \rangle ds) = A_0 \exp(\mu y), 
\end{eqnarray}
with the normalization factor $A_0=\sqrt{\mu/e^{(2 \mu L-1)}} $. There is no zero mode company for $g_{0^+}$ but each of the KK modes has its couple and fill the spinor states of the Dirac spinor. 
The solution for the even KK modes is 
\begin{eqnarray}
\label{Eq:waveven}
g_{n^+} &=& \sqrt{\frac{1}{L}} \Big[ \frac{k_{n^+}}{m_{n^+}} \cos(k_{n^+} y) + \frac{\mu}{m_{n^+}}\sin(k_{n^+} y)\Big], \nonumber \\
f_{n^+} &=& - \sqrt{\frac{1}{L}} \sin(k_{n^+} y),
\end{eqnarray}
where the KK mass $m_{n^+} = \sqrt{\mu^2 + k_{n^+}^2}$ and $k_{n^+} = {n \pi }/{L}$ ($n \in \mathcal{Z}$).

To avoid the very light 1st KK quark, we choose $\mu > 0$ and the zero mode is quasi-localized at the boundary $y = \pm L$. For odd KK modes, the solution is
\begin{eqnarray}
\label{Eq:waveodd}
g_{n^-} &=& - \sqrt{\frac{1}{L}} \sin(k_{n^-} y), \nonumber \\
f_{n^-} &=& \sqrt{\frac{1}{L}} \Big[ \frac{k_{n^-}}{m_{n^-}} \cos(k_{n^-} y) + \frac{\mu}{m_{n^-}}\sin(k_{n^-} y)\Big],
\end{eqnarray}
where the KK mass $m_n = \sqrt{\mu^2 + k_n^2}$ and $k_n$ is the $n$-th solution of the equation 
\bea
\label{Eq:odd}
k_{n^-} = -\mu \tan(k_{n^-} L) \ .
\eea When $\mu$ increases from 0 to $+ \infty$, $k_{n^-}$ increases from $(n-1/2) \pi / L$ to $n \pi / L$. In this case, in the limit of $\mu \rightarrow + \infty$, all KK modes could be decoupled.


\end{document}